\documentclass[11pt,reqno,a4paper]{amsart}

\usepackage{amsmath}
\usepackage{enumerate}
\usepackage{amssymb}
\usepackage{amscd}
\usepackage{amsthm}
\usepackage{amsfonts}
\usepackage{comment}
\usepackage{graphicx}
\usepackage{hyperref}
\usepackage[matrix, arrow, curve]{xy}
\usepackage{etoolbox}
\usepackage{mdframed}
\usepackage{setspace}
\usepackage{mathrsfs}  
\DeclareMathSymbol{\shortminus}{\mathbin}{AMSa}{"39}

\numberwithin{equation}{section}

\patchcmd{\subsection}{-.5em}{.5em}{}{}
\patchcmd{\subsubsection}{-.5em}{.5em}{}{}

\newcommand{\GL}{\operatorname{GL}}

\newcommand{\Aut}{\operatorname{Aut}}

\newcommand{\cB}{\mathcal{B}}
\newcommand{\cC}{\mathcal{C}}

\newcommand{\cE}{\mathcal{E}}
\newcommand{\cF}{\mathcal{F}}

\newcommand{\cH}{\mathcal{H}}

\newcommand{\cL}{\mathcal{L}}

\newcommand{\cO}{\mathcal{O}}
\newcommand{\cP}{\mathcal{P}}

\newcommand{\cT}{\mathcal{T}}

\newcommand{\bC}{\mathbb{C}}

\newcommand{\bE}{\mathbb{E}}

\newcommand{\bN}{\mathbb{N}}

\newcommand{\bP}{\mathbb{P}}
\newcommand{\bQ}{\mathbb{Q}}
\newcommand{\bR}{\mathbb{R}}

\newcommand{\bT}{\mathbb{T}}

\newcommand{\bZ}{\mathbb{Z}}

\newcommand{\R}{\mathbb{R}}
\newcommand{\C}{\mathbb{C}}

\newcommand{\ra}{\rightarrow}

\newcommand{\qand}{\quad \textrm{and} \quad}

\newcommand\subsetsim{\mathrel{%
\ooalign{\raise0.2ex\hbox{$\subset$}\cr\hidewidth\raise-0.8ex\hbox{\scalebox{0.9}{$\sim$}}\hidewidth\cr}}}
\newcommand{\eps}{\varepsilon}

\DeclareMathOperator{\pr}{pr}

\DeclareMathOperator{\supp}{supp}

\DeclareMathOperator{\Per}{Per}
\DeclareMathOperator{\Var}{Var}
\DeclareMathOperator{\Cov}{Cov}
\DeclareMathOperator{\Vol}{Vol}
\newcommand{\Q}{\mathbb Q}
\newcommand{\Z}{\mathbb Z}

\renewcommand{\epsilon}{\varepsilon}
\newcommand{\dd}{\mathrm{d}}
\DeclareMathOperator{\Proj}{Proj}

\DeclareMathOperator{\ANV}{ANV}
\DeclareMathOperator{\covol}{covol}

\theoremstyle{theorem}
\newtheorem{theorem}{Theorem}[section]
\newtheorem{corollary}[theorem]{Corollary}
\newtheorem{proposition}[theorem]{Proposition}
\newtheorem{lemma}[theorem]{Lemma}

\theoremstyle{definition}
\newtheorem{definition}[theorem]{Definition}
\newtheorem{assumption}[theorem]{Assumption}

\newtheorem{construction}[theorem]{Construction}
\newtheorem{remark}[theorem]{Remark}
\newtheorem{example}[theorem]{Example}
\newtheorem{problem}[theorem]{Problem}
\newtheorem{question}[theorem]{Question}

\renewcommand{\phi}{\varphi}

\begin{document}

\title{Hyperuniformity and non-hyperuniformity of quasicrystals}

\author{Michael Bj\"orklund}
\address{Department of Mathematics, Chalmers, Gothenburg, Sweden}
\email{micbjo@chalmers.se}
\thanks{}

\author{Tobias Hartnick}
\address{Institut f\"ur Algebra und Geometrie, KIT, Karlsruhe, Germany}
\curraddr{}
\email{tobias.hartnick@kit.edu}
\thanks{}




\date{}

\dedicatory{}

\maketitle

\begin{abstract} 
We develop a general framework to study hyperuniformity of various mathematical models of quasicrystals. Using this framework we provide examples of non-hyperuniform quasicrystals which unlike previous examples are not limit-quasiperiodic. Some of these examples are even anti-hyperuniform or have a positive asymptotic number variance. On the other hand we establish hyperuniformity for a large class of mathematical quasicrystals in Euclidean spaces of arbitrary dimension. For certain models of quasicrystals we moreover establish that hyperuniformity holds for a generic choice of the underlying parameters. For quasicrystals arising from the cut-and-project method we conclude that their hyperuniformity depends on subtle diophantine properties of the underlying lattice and window and is by no means automatic.
\end{abstract}

\section{Introduction}
In an influential article in 2003 \cite{TS2}, Torquato and Stillinger initiated a systematic study of point processes with reduced long-wavelength density fluctuations under the name of \emph{hyperuniformity}\footnote{In some parts of the literature the term \emph{superhomogeneity} is used.}. Over the last 20 years hyperuniform point processes have been studied intensely from a mathematical point of view and found applications in physics, materials science, chemistry, engineering and even biology; we refer to the survey of Torquato \cite{Torquato} for an extensive bibliography. In particular, the question of hyperuniformity has been investigated for various mathematical models of quasicrystals \cite{BG, BG2, OSST2, OSST}.

This article investigates hyperuniformity (and various related properties) for a class of pure point diffractive jammed hard-core point processes known as \emph{cut-and-project processes}, which in dimensions $2$ and $3$ provides one of the most widely accepted mathematical models for quasicrystals. It was observed by O\u{g}uz, Socolar, Steinhardt and Torquato \cite{OSST2} (see also related work by Baake and Grimm \cite{BG}) that many one-dimensional cut-and-project processes are hyperuniform. On the other hand, it was already pointed out in \cite{OSST} by the same authors that not all cut-and-project processes are hyperuniform. However, the previously known counterexamples seem to be of a very special form, namely \emph{limit-quasiperiodic}.

In the present article we provide examples of cut-and-project processes which are \emph{not} hyperuniform (in a very strong quantitative sense) without being limit-quasiperiodic. On the other hand, we also establish hyperuniformity for large classes of cut-and-project processes (in arbitrary dimension) with spherical Euclidean (or more generally, Fourier smooth) windows, and for such cut-and-project processes we even prove that hyperuniformity holds for \emph{generic} choices of the underlying lattice. The main upshot of our discussion will be that hyperuniformity of quasicrystals depends on subtle diophantine properties of the underlying data and is by no means automatic.

\subsection{On the definition of hyperuniformity}

Denote by $B_R$ the Euclidean ball of radius $R$ around $0$ in $\R^d$. A locally-square integrable invariant point process $\Lambda$ in $\R^d$ is called \emph{geometrically hyperuniform} (with respect to Euclidean balls) if its \emph{asymptotic number variance}
\[
\ANV(\Lambda) := \lim_{R \to \infty} \frac{\mathrm{Var}(\#(\Lambda \cap B_R))}{\Vol_d(B_R)}
\]
exists and is equal to $0$. Here the denominator $\Vol_d(B_R)$ can be interpreted as the number variance of a suitably normalized Poisson process $\Lambda_{\mathrm{Pois}}$, and hence geometric hyperuniformity corresponds to a sub-Poissonian number variance for large balls. This property can also be expressed spectrally: Given a bounded measurable function $f$ on $\R^d$ with bounded support, we denote by $\cP_\Lambda f(\omega) = \sum_{x \in \Lambda_\omega} f(x)$ the corresponding linear statistic. Then there exists a unique positive-definite signed measure $\eta_\Lambda$ on $\R^d$ such that
\[
\eta_\Lambda(f \ast f^*) = \Var(\cP_\Lambda f),
\]
and we denote by $\widehat{\eta}_\Lambda$ its Fourier transform\footnote{$\widehat{\eta}_\Lambda$ can be obtained from the diffraction measure of $\Lambda$ by removing the atom at $0$.
If $\widehat{\eta}_\Lambda$ has a density with respect to Lebesgue measure, then this density is called the \emph{structure factor} of the process; however, the processes considered in this article will have pure point diffraction, hence their structure factor is not defined.}. For example, if $\Lambda_{\mathrm{Pois}}$ is a Poisson process on $\R^d$, then up to scaling we have $\widehat{\eta}_{\Lambda_{\mathrm{Pois}}} = \mathrm{Vol}_d$. We then say that $\Lambda$ is \emph{spectrally hyperuniform} (with respect to Euclidean balls) if $\widehat{\eta}_\Lambda$ decays near $0$ faster than a Poisson process, i.e.\ 
\[
 \lim_{\eps \to 0} \frac{\widehat{\eta}_\Lambda(B_\eps)}{\widehat{\eta}_{\Lambda_{\mathrm Pois}}(B_\eps)} = \lim_{\eps \to 0} \frac{\widehat{\eta}_\Lambda(B_\eps)}{\Vol_d(B_\eps)}= 0.
\]
The following result was previously established in various special cases (see e.g.\ \cite[Proposition 2.2]{C}); we provide a proof of a more general statement in Theorem \ref{TheoremSvsG} below. 
\begin{theorem}[Geometric vs.\ spectral hyperuniformity]\label{EquivHU} A locally-square integrable invariant point process $\Lambda$ in $\R^d$ is geometrically hyperuniform if and only if it is spectrally hyperuniform.
\end{theorem}
We emphasize that Theorem \ref{EquivHU} only holds with respect to \emph{Euclidean} balls. For general balls, spectral hyperuniformity implies geometric hyperuniformity, but as observed by Kim and Torquato \cite{KT} the standard lattice $\Z^2$ in $\R^2$ is spectrally hyperuniform, but not geometrically hyperuniform with respect to balls in the $\ell^\infty$-metric (cf.\ \cite[Section 2.1]{C}).



\subsection{Meyerian point processes}

Meyerian point processes are a class of point processes in Euclidean space which in dimensions $2$ and $3$ provide mathematical models for quasicrystals. By definition, a subset $\Lambda \subset \R^d$ is called a \emph{Meyer set} if it is relatively dense (i.e. $\Lambda + K = \R^d$ for some compact subset $K \subset \R^d$) and if $\Lambda-\Lambda$ is uniformly discrete. By a \emph{Meyerian point process} we shall mean an ergodic point process in $\R^d$ consisting of random Meyer sets. Note that Meyer sets are in particular relatively dense and uniformly discrete, hence in probabilistic language Meyerian point processes are examples of \emph{jammed hard-core processes}. In the present article we are mostly interested in a specific class of Meyerian point processes with pure point diffraction which arise from the cut-and-project construction.

To define this class of processes, let $G$ and $H$ be locally compact abelian groups and let $\Gamma$ be a lattice in $G \times H$ which projects injectively to $G$ and densely to $H$, and let $W \subset H$ be a relatively compact subset with dense interior. If $G = \R^d$, then any translate of the cut-and-project set
\[
\Lambda(G,H,\Gamma,W) := \pr_G(\Gamma \cap (G \times W)) \subset G
\]
is a Meyer set, and we thus obtain a Meyerian point process $\Lambda$ parametrized by $\Omega := \Gamma \backslash (G \times H)$ with its unique $G$-invariant probability measure by setting
\[
\Lambda_{[g,h]} := \Lambda(G, H, \Gamma, Wh^{-1})g.
\]
We refer to this point process as a \emph{cut-and-project process with parameters $(G, H, \Gamma, W)$}. The group $H$ is called the \emph{internal space} of $\Lambda$. 

As we will see, in certain situations the question of hyperuniformity of cut-and-project processes can be related to \emph{diophantine} properties of the underlying lattice.

\subsection{Non-hyperuniform cut-and-project processes}
It is well-known that cut-and-project processes in $\R^d$ with a totally disconnected internal space need not be hyperuniform. In \cite[Section B]{OSST},  the authors investigate hyperuniformity for a large class of one-dimensional limit-quasiperiodic quasicrystals.  Among other things, they produce a limit-quasiperiodic random quasicrystal $\Lambda$ and a lacunary sequence $(R_n)$ of radii such that the limit
\[
c = \lim_{n \ra \infty}  \frac{\mathrm{Var}(\#(\Lambda \cap B_{R_n}))}{\Vol_d(B_{R_n})}
\]
exists and is positive. This shows that $\Lambda$ is not hyperuniform, but it does not imply that the limit defining $\ANV(\Lambda)$ exists. The connection to cut-and-project processes is through the observation that certain random limit quasi-periodic quasicrystals can be realized as cut-and-project-processes whose internal space has a totally disconnected factor \cite{BMS}. To the best of our knowledge, the following two problems have remained open so far: 
\begin{problem}\label{PI} Does there exist a non-hyperuniform cut-and-project process with connected internal space?
\end{problem}
\begin{problem}\label{PII} Does there exist a Meyerian point process with positive asymptotic number variance?
\end{problem}
In this article, we answer both questions in the affirmative. The following consequence of Theorem \ref{Thm_NonHypCP} below solves Problem \ref{PI}.
\begin{theorem}[Cut-and-project processes need not be hyperuniform]\label{AntiHUIntro}
For every exponent $\alpha > 0$ there exists a cut-and-project process with parameters $(\R, \R, \Gamma, W)$ and centered diffraction $\widehat{\eta}$
such that $W \subset \R$ is an interval and \[
\varlimsup_{\eps \to 0} \frac{\widehat{\eta}(B_\eps)}{\eps^\alpha} = \infty.
\]
\end{theorem}
Theorem \ref{AntiHUIntro} says that, even in one dimension (and one ``internal'' dimension), there exist cut-and-project processes for which the diffraction of small balls of radius $\eps$ converges to $0$ more slowly than any positive power of $\eps$ along \emph{some} sequence of radii. In the proof of Theorem \ref{AntiHUIntro} these radii will be chosen carefully according to certain diophantine properties of the lattice $\Gamma$ and window $W$; with our method we cannot control the behaviour along arbitrary sequences of radii.

In order to solve Problem \ref{PII}, we thus consider a different class of examples. Using mixing dynamical systems, we construct in Section \ref{SecMeyerian} below a class of Meyerian point process $\Lambda_o$ in $\Z$ which are $2$-syndetic in the sense that two translates of $(\Lambda_o)_\omega$ cover $\Z$ for every $\omega$. Via suspension, we can extend any such process $\Lambda_o$ to a Meyerian point process in $\Lambda$ in $\R$, which is a $2$-syndetic subset of a random translate of $\Z$ in $\R$. The following theorem is then a consequence of Theorem \ref{PosANV} below; here a subset of $\Z$ is called \emph{$2$-syndetic} if two of its translates cover $\Z$.
\begin{theorem}[Meyerian point processes may have positive asymptotic number variance]\label{PosANVIntro}
There exists a Meyerian point process $\Lambda$ in $\R$, which is $2$-syndetic in a random translate of $\Z$ and for which the asymptotic number variance $\ANV(\Lambda)$ exists and is strictly positive.
\end{theorem}

\subsection{Hyperuniform cut-and-project processes}
An important class of lattices in products of Euclidean spaces is the class of \emph{arithmetic lattices} (see Example \ref{ExArithmetic} below). For the associated cut-and-project processes with spherical windows Corollary \ref{HUArithmetic} below implies the following:
\begin{theorem}[Arithmetic cut-and-project processes are hyperuniform]\label{ArithmeticIntro} Let $\Gamma < \R^{d_1+d_2}$ be an arithmetic lattice and let $W \subset \R^{d_2}$ be a Euclidean ball. Then the cut-and-project process with parameters $(\R^{d_1}, \R^{d_2}, \Gamma, W)$ is hyperuniform and its centered diffraction $\widehat{\eta}$ satisfies
\[
\frac{\widehat{\eta}(B_\eps)}{\eps^{d_1}} \ll \eps^{\frac{d_1}{d_2}}.
\]
\end{theorem}
Given that there exist both hyperuniform and non-hyperuniform quasicrystals, the question becomes relevant which of the two behaviours is ``generic''. To answer this question we note that there is a unique $\mathrm{GL}_{d_1+d_2}(\R)$-invariant measure class on the space of all lattices in $\R^{d_1+d_2}$, which allows us to talk about generic lattices with respect to this measure class. Then the following is a special case of Theorem \ref{GenericHyperuniformity} below:
\begin{theorem}[Generic quasicrystals are hyperuniform]\label{GenericIntro} Let $W$ be a Euclidean ball. Then for every $\delta > 0$ there exists a conull set of lattices $\Gamma < \R^{d_1+d_2}$ such that the cut-and-project process with parameters $(\R^{d_1}, \R^{d_2}, \Gamma, W)$ is hyperuniform and its diffraction $\widehat{\eta}$ satisfies
\[
\frac{\widehat{\eta}({B}_\eps)}{\eps^{d_1}} \ll_\delta \eps^\frac{d_1(1-\delta)}{d_2+\delta}.
\]
\end{theorem}
Thus, as far as hyperuniformity is concerned, generic lattices are only marginally worse than arithmetic ones. To keep the formulation simple, we have formulated Theorem \ref{ArithmeticIntro} and Theorem \ref{GenericIntro} only for Euclidean balls. For cut-and-project processes with more general windows the decay rate of the diffraction at $0$ depends in an explicit way on the Fourier decay of the window. One thus obtains hyperuniformity as soon as the window has sufficient Fourier decay (see Corollary \ref{HUArithmetic} and Theorem \ref{GenericHyperuniformity}). Both Theorem \ref{ArithmeticIntro} and Theorem \ref{GenericIntro} (and their generalizations to windows with sufficient Fourier decay) are based on a hyperuniformity criterion which uses the following notion of \emph{repellence} of lattices in products.
\begin{definition}\label{DefRep}
A lattice $\Gamma < \bR^{d_1} \times \bR^{d_2}$ is 
\emph{$\beta$-repellent on the right} if
there exists $\eps_o > 0$ such that for every $(\gamma_1,\gamma_2) \in \Gamma \setminus \{(0,0\}$ and all $\eps < \eps_o$,
\[
\|\gamma_1\|_\infty < \eps \implies \|\gamma_2\|_\infty \geq \eps^{-\beta}.
\]  
\end{definition}
With this notion, we have the following sufficient condition for hyperuniformity, which is a special case of Theorem \ref{SufficientCondition}  below:
\begin{theorem}[Sufficient condition for hyperuniformity]\label{SufficientConditionIntro}  Let $W$ be a Euclidean ball and let $\widehat{\eta}$ denote the diffraction of the cut-and-project process with parameters $(\R^{d_1}, \R^{d_2}, \Gamma, W)$. If the dual lattice $\Gamma^{\perp}$ of $\Gamma$ is $\beta$-repellent on the right for some $\beta > 0$, then for all sufficiently small $\varepsilon > 0$ we have
\[
 \frac{\widehat{\eta}_p({B}_\eps)}{{\Vol}_{d_1}(B_\eps)} \ll \eps^{{\beta(d_2+1)} - d_1}
\]
In particular, the quasicrystal is hyperuniform provided that ${\beta} > \frac{d_1}{{d_2 + 1}}$.
\end{theorem}
Again, the theorem applies also to more general windows, but then the exponent will depend on the Fourier decay of the window (see Theorem \ref{SufficientCondition}).
Notably, the lattices appearing in Theorem \ref{ArithmeticIntro} and Theorem \ref{GenericIntro} satisfy the repellence condition of Theorem \ref{SufficientConditionIntro}. On the other hand, our construction of non-hyperuniform cut-and-project processes for Theorem \ref{AntiHUIntro} starts from a lattice, whose dual lattice is not sufficiently repellent for the criterion to apply. However, since repellence is only sufficient and not necessary for hyperuniformity, additional work is required to produce non-hyperuniform cut-and-project processes, and in particular the choice of window plays a crucial role.

\subsection{Number rigid cut-and-project processes}
We recall that a locally square-integrable invariant point process $\Lambda$ in $\R^d$ is called \emph{number rigid} if for every Borel set $B \subset \R^d$ the number $\# (B \cap \Lambda)$ of points in $B$ depends almost surely only on $\Lambda|_{B^c}$. While number rigidity is neither implied by nor implies hyperuniformity, the two properties are nevertheless related. In Lemma \ref{NRCrit} below we adapt an argument of Ghosh and Peres \cite{GP} to establish the following spectral criterion for number rigidity in the spirit of spectral hyperuniformity.
\begin{lemma}[Spectral criterion for number rigidity]\label{NRCritIntro}
A locally-square integrable invariant point process $\Lambda$ in $\R^d$ is number rigid, provided there is a sequence $\eps_n \searrow 0$ such that the centered diffraction of $p$ satisfies
\[
\widehat{\eta}_p(B_{\eps_n}) \ll \eps_n^{2d+\delta}.
\]
\end{lemma}
Compared to spectral hyperuniformity we require a much stronger decay with exponent $2d+\delta$ rather than $d+\delta$; on the other hand, this decay is only required along one specific sequence. The decay required to apply Lemma \ref{NRCritIntro} is much stronger than what we establish for generic (or even uniform) cut-and-project-processes with spherical window above. However, it is well-known that one can obtain stronger decay by making very specific choices of windows (often related to self-similarity phenomena). For example, by a result of Baake and Grimm \cite[Proposition 3.1]{BG},  the Fibonacci cut-and-process process ($d=1$) satisfies the condition of the lemma with $\delta = 2$. We deduce:
\begin{corollary}[Number rigid cut-and-project processes] The Fibonacci cut-and-project process is number rigid.\qed
\end{corollary}
At least with known methods, it is not possible to verify the criterion from Lemma \ref{NRCritIntro} for cut-and-project processes which are obtained from the Fibonacci process by modifying the window (see the discussion in \cite[Section 7]{BG2}). However, this does not mean that such processes are not number rigid. Neither is the criterion from Lemma \ref{NRCritIntro} necessary, nor is it possible with current techniques (including ours) to provide lower bounded for the centered diffraction of a cut-and-project process along \emph{all} sequences of radii. The following problem therefore remains open:
\begin{question} Do there exist cut-and-project processes which are not number rigid?
\end{question}

\subsection{Stealthy cut-and-project processes}
A property of locally square-integrable point processes which is stronger than both hyperuniformity and number rigidity is vanishing of the centered diffraction in a neighbourhood of $0$. For example, this property holds for periodic point processes. More generally, a locally square-integrable point process $\Lambda$ is called \emph{stealthy} if its centered diffraction $\widehat{\eta}_p$ vanishes identically on \emph{some} open set. A recent, and quite surprising, result \cite[Theorem 3]{BSW} says that every stealthy point process in $\Z$ is periodic. Stealthy point processes (and random measures) on $\bR^d$ have been thoroughly investigated in \cite{STZ} and \cite{GL2}. To construct stealthy cut-and-project processes we have to pass from the real to the $p$-adic setting. For this we note that cut-and-project processes and their centered diffraction measures can also be considered in other locally compact abelian groups, and diffraction measures can be defined as in the real case. In the $p$-adic context, we establish the following result (cf.\ Corollary \ref{stealth}):

\begin{theorem}[Stealthy $p$-adic cut-and-project processes]\label{stealthIntro} There exists a stealthy cut-and-project process in $\Q_p$ (with $\bR$ as internal space) for every prime $p$.
\end{theorem}

\subsection{Organization of the article}
This article is organized as follows: In a preliminary Section \ref{SecPrel} we collect various basic facts concerning autocorrelation and diffraction measures of invariant point processes (or, more generally, invariant random measures). In Section \ref{SecHU} we discuss the definition of hyperuniformity and establish Theorem \ref{EquivHU}. Sections \ref{Counter} -- \ref{SecMeyerian} form the core of the article and discuss examples of hyperuniformand non-hyperuniform quasicrystals. In Section \ref{Counter} we establish Theorem \ref{AntiHUIntro}, in Section \ref{Criterion} we establish Theorem \ref{SufficientConditionIntro} and derive Theorem \ref{ArithmeticIntro} and Theorem \ref{GenericIntro} and in Section \ref{SecMeyerian} we establish Theorem \ref{PosANVIntro}. The remainder of the article then discusses various related properties: In Section \ref{SecNR} and Section \ref{SecStealth} we construct cut-and-project processes which are number rigid and stealthy respectively, thereby proving Lemma \ref{NRCritIntro} and Theorem \ref{stealthIntro}.

For the convenience of the reader we include two appendices. Appendix \ref{DiffFormulas} explains the well-known diffraction formulas for Poisson processes and cut-and-project processes, whereas Appendix \ref{AppTrans} contains an introduction to the general formalism of transverse point processes, a wide class of hard-core point processes recently introduced in \cite{BHK}, which contains all of the point processes considered in this article.

\subsection{Notations and conventions}
The letter $G$ will always be reserved for a locally compact second countable (lcsc) group, which is always assumed to be unimodular and often assume to be abelian.
We will denote by $\cL^\infty_c(G)$ the space of bounded complex-valued Borel functions on $G$ which vanish outside a compact subset of $G$ and by $C_c(G)$ its subspace of complex-valued compactly supported continuous functions. Dually we denote by $M(G)$ the space of signed Radon measures on $G$. Given a space $\cF$ of functions or measures on which $G$ acts, we denote by $\cF^G \subset \cF$ the subspace of $G$-invariants.

All probability spaces considered in this article are standard, i.e.\ the underlying Borel space is standard. Given a probability space $(\Omega, \cB, \cP)$ we denote by $L^2(\Omega, \bP)$ the corresponding $L^2$-space and by $L^2_o(\Omega, \bP)$ the orthogonal complement of the constant functions.

Given non-negative real-valued functions $f$ and $g$ on some space $X$ we write $f(x) \ll g(x)$ if there exists a constant $C>0$ such that $f(x) \leq C g(x)$ for all $x \in X$. We write $f(x) \ll_\alpha g(x)$ to indicate that $C$ depends on some parameter $\alpha$. If these inequalities only hold in some asymptotic sense, then we use the usual Landau notation $o$ and $O$, where indices like $O_\alpha$ indicate again parameters on which the constants depend.

\section{Preliminaries on autocorrelation and diffraction}\label{SecPrel}

\subsection{Autocorrelation of invariant random measures}
In this article we are mainly interested in certain invariant simple point processes in Euclidean space $\R^d$ of arbitrary dimension $d$. However, it is convenient to define some of the basics notions related to point processes in their natural generality.

Thus let $G$ be a unimodular lcsc group with Haar measure $m_G$ acting probability-measure preservingly (pmp) on a probability space $(\Omega, \mathcal F, \mathbb{P})$. Every $f \in \cL_c^\infty(G)$ defines a function \[\cP f: M(G) \to \C, \quad p \mapsto p(f)\] on the space of signed Radon measures, called the associated \emph{linear statistic}, and we equip $M(G)$ with the smallest $\sigma$-algebra $\cB_{M(G)}$ for which all of these linear statistics are measurable. Then a $G$-equivariant measurable map
\[
p: \Omega \to M(G), \quad \omega \mapsto p_\omega
\]
is called an \emph{invariant\footnote{In the probabilistic literature the term ``stationary'' is sometimes used instead of invariant.} random measure} on $G$ with \emph{distribution} $\mu_p := p_*\bP \in \mathrm{Prob}(M(G))^G$.  We say that $p$ is \emph{ergodic} if $\bP$ is ergodic and refer to the moments of $\mu_p$ as the moments of $p$.

An invariant random measure $p$ is called an \emph{invariant point process} if $p_\omega$ is $\mathbb{P}$-almost surely supported in a locally finite subset of $G$ and \emph{simple} if moreover all atoms of $p_\omega$ have size $1$. In this case we can identify $p_\omega$ with its support and thereby think of $p$ as an invariant random locally finite subset of $G$. 
\begin{example}[Cut-and-project processes]\label{ExCuP} Let $G$ and $H$ be unimodular locally compact second countable (lcsc) groups, let $\Gamma < G \times H$ be a a lattice which projects injectively to $G$ and densely to $H$ and let $W \subset H$ be a relatively compact subset with dense interior. We define $\Omega := \Gamma\backslash(G \times H)$ and denote by $\bP$ the unique $G$-invariant probability measure on $\Omega$ (cf. \cite{BHP1}). One can show that the map
\[
p = p(G, H, \Gamma, W): (\Omega, \bP) \to M(G), \quad \Gamma(g,h) \mapsto \delta_{\Lambda(G, H, \Gamma, Wh^{-1})g}
\]
is a well-defined Borel map, hence defines a simple hard-core point process. This is a special case of a general construction discussed in more detail in Appendix \ref{AppTrans}. Generalizing the definition from the introduction, we refer to $p(G, H, \Gamma, W)$ as a \emph{cut-and-project process} with parameters $(G, H, \Gamma, W)$. If $G = \R^{d_1}$ and $H = \R^{d_2}$, then the space $\Omega$ is a torus of dimension $d_1 + d_2$, and the $G$-orbits form a foliation of this torus by $d_1$-dimensional leaves, which are dense embeddings of $\R^{d_1}$.
\end{example}
Note that cut-and-project processes automatically satisfy the following assumption.
\begin{assumption} All invariant random measures in this article are assumed to be \emph{locally square-integrable} in the sense that
\[\bE[p_\omega(B)^2] < \infty \quad (B \subset G\; \text{bounded Borel set}).\]
\end{assumption}
From now on $p$ will always denote a locally square-integrable invariant random measure with distribution $\mu_p$.
This assumption ensures that the first two moment measures  $M^1_p \in M(G)^G$ and  $M^2_p \in M(G \times G)^G$  of $p$ exist. By definition these
two moment measures are then given by
\[
M^1_p(A) := \bE[p_\omega(A)] \qand M^2_p(B) := \bE[(p_\omega \otimes p_\omega)(B)]
\] 
for bounded Borel sets $A \subset G$ and $B \subset G \times G$. Since $M^1_p$ is invariant, there exists a constant $i_p$, called the \emph{intensity} of the random measure $p$ with respect to $m_G$, such that
\begin{equation}\label{Intensity}
M^1_p = i(p) \cdot m_G
\end{equation}
Similarly, since $M^2_p$ is $G$-invariant, under the identification $(G \times G)/\Delta(G)\to G$ given by $[(g_1,g_2)] \mapsto g_1g_2^{-1}$ it corresponds to a Radon measure ${\eta}^+_p$ on $G$ called its autocorrelation (cf.\ \cite{BHP1}). This measure satisfies
\[
\eta^+_p(f_1 \ast \check f_2) = M^2_p(f_1 \otimes f_2) \quad (f_1, f_2 \in \cL^\infty_c(G)),
\]
and if $\rho$ is a non-negative Borel function on $G$ with bounded support, normalized to $m_G(\rho) = 1$, then for every bounded Borel function $f$ on $G$ with bounded support we have
\begin{equation}\label{AutoCorrelationExplicit}
\eta^+_p(f) = \int_{G \times G} f(g_1 g_2^{-1})\rho(g_2) \, \dd M^2_p(g_1, g_2),
\end{equation}
independently of $\rho$. The measure  ${\eta}^+_p$ is positive-definite, since for $f_1, f_2, f \in C_c(G)$ we have
\begin{equation}
 \eta^+_p(f_1 \ast f_2^*) = \langle p_\omega(f_1), p_\omega(f_2) \rangle_{L^2(\Omega, \bP)}, \quad \text{and hence} \quad \eta^+_p(f \ast f^*) = \| p_\omega(f)\|^2 _{L^2(\Omega, \bP)} \geq 0.
\end{equation}
It turns out that the measure $\eta^+_p$ is closely related to the variance $\Var_p$ of the process $p$, where, by definition, $\Var_p(A) = \Var[\cP \chi_A]$ is just the variance of the real-valued random variable $\omega \mapsto p_\omega(A)$ for any bounded Borel set $A \subset G$. The connection becomes apparent if we define a signed Radon measure on $G$ by the formula
\begin{equation}\label{etap}
 \eta_p(f) :=  \eta_p^+(f ) - i(p)^2 \cdot m_G(f) \quad (f \in \cL_c^\infty(G)).
\end{equation}
\begin{proposition}[Variance vs.\ autocorrelation] The signed measure $\eta_p$ satisfies
\[
 \eta_p(f \ast f^*) = \Var_p(f) \quad (f \in \cL^\infty_c(G))
\]
and is uniquely determined by this property. In particular, it is positive-definite.
\end{proposition}
\begin{proof} It is convenient to introduce the \emph{covariance} of $p$ as the signed measure given by the centered second moment
\begin{eqnarray*}
\Cov_p(f_1 \otimes f_2) &=& \bE\left[\left(p_\omega(f_1)- \bE[p_\omega(f_1)]\right) \left(p_\omega(f_2) - \bE[p_\omega(f_2)]\right) \right]\\ &=& M^2_p(f_1 \otimes f_2) - M_p^1(f_1) M_p^1(f_2)  \quad (f_1, f_2 \in \cL^\infty_c(G)).
\end{eqnarray*}
Since this is $G$-invariant, it corresponds to a signed measure $\eta_p$ on $G$ such that $\eta_p(f_1 \ast \check f_2) = \Cov_p(f_1 \otimes f_2)$, and we claim that this signed measure satisfies formula \eqref{etap}. Indeed, for all $f,g \in \cL_c^\infty(G)$ we have
\[
 \eta_p(f \ast g^*)=  M^2_p(f \otimes \bar g) - M_p^1(f) M_p^1(\overline g) = \eta_p^+(f \ast g^*) -  i(p)^2 \cdot m_G(f) m_G(\overline g).
\]
Now let $(U_n)$ be a nested sequence of compact identity neighbourhoods converging to $\{e\}$ and let $\rho_n$ be a non-negative function supported in $U_n$ and normalized to integral $1$. Then $(\rho_n)$ is an approximate identity and hence
\[
\eta_p(f) = \lim_{n \to \infty} \eta_p(f \ast \rho_n^*) = \lim_{n \to \infty}\left( \eta_p^+(f \ast \rho_n^*) -  i(p)^2 \cdot m_G(f) m_G(\overline \rho_n)\right) = \eta_p^+(f) - i(p)^2 \cdot m_G(f).
\]
This proves \eqref{etap}, and the latter determines $\eta_p^+$ uniquely. Finally,
\[
\eta_p(f \ast f^*) = \Cov_p(f \otimes \bar f) =  \bE\left[\left| p_\omega(f)- \bE[p_\omega(f)]\right|^2 \right] = \Var(p_\omega(f)) = \Var_p(f).\qedhere
\]
\end{proof}
Both of the closely related (signed) measures $\eta^+_p$ and $\eta_p$ are sometimes called the \emph{autocorrelation measure} of $p$ (or $\mu_p$) in the literature. We will reserve this term for the measure $\eta^+_p$ and refer to $\eta_p$ as the \emph{centered autocorrelation measure}. 

\subsection{Diffraction of invariant random measures on LCA groups}
We now assume that $G$ is an \emph{abelian} lcsc group and denote by $\widehat{G}$ its Pontryagin dual. We then normalize Haar measures $m_G$ and $m_{\widehat{G}}$ such that Fourier inversion becomes
\begin{equation}\label{DualMeasures}
f(x) = \int_{\widehat{G}} \widehat{f}(\chi) \chi(x) \, \dd m_{\widehat{G}}(x), \quad \text{where} \quad \widehat{f}(\chi) = \int_G f(x) \overline{\chi(x)} \, \dd m_G(x),
\end{equation}
for all sufficiently regular functions on $G$. 
Given a positive definite signed Radon measure $\eta$ on $G$ we denote by $\widehat{\eta}$ its Fourier transform (cf.\ \cite[Theorem 4.5]{BF}), which is a (positive) Radon measure on $\widehat{G}$.

From now on let $p$ be a locally square-integrable invariant random measure on $\R^d$ with autocorrelation $\eta^+_p$ and centered autocorrelation $\eta_p$. Since these are positive-definite, we can take their respective Fourier transforms  $\widehat{\eta}^+_p$ and $\widehat{\eta}_p$.
\begin{definition}
$\widehat{\eta}^+_p$ is called the \emph{diffraction} of $p$, and $\widehat{\eta}_p$ is called the \emph{centered diffraction} of $p$. 
\end{definition}
We note that by definition
\begin{equation}\label{Diffraction}
\widehat{\eta}^+_p(|\widehat{f}|^2) = \eta^+_p(f * f^*) =  \|p_\omega(f)\|_{L^2(\Omega, \bP)}^2 
\qand  \widehat{\eta}_p(|\widehat{f}|^2) = \eta_p(f * f^*) =\Var_p(f),
\end{equation}
for all sufficiently regular functions $f$,  including all compactly supported continuous functions.   A straightforward approximation argument then shows that \eqref{Diffraction} holds for all $f \in \cL^\infty_c(G)$.  

\vspace{0.1cm}

\begin{proposition}[Diffraction vs.\ centered diffraction]\label{RemoveAtom}
If $p$ is ergodic, then the diffraction and the centered diffraction are related by the formula \[\widehat{\eta}_p =  \widehat{\eta}^+_p -  \widehat{\eta}^+_p(\{1\}) \cdot \delta_1.\]
\end{proposition}
Here, $1 \in \widehat{G}$ denotes the trivial character. Note that, by \eqref{etap} we have for all $f \in \cL^\infty_c(G)$,
\[
  \widehat{\eta}_p(|\widehat{f}|^2) = \eta_p(f \ast f^*) = \eta_p^+(f \ast f^*) - i(p)^2 \cdot m_G(f \ast f^*) = \widehat{\eta}^+_p(|\widehat{f}|^2) - i(p)^2 \cdot |\widehat{f}|^2(0).
\]
Thus Proposition \ref{RemoveAtom} reduces to the following lemma:
\begin{lemma}[Intensity formula]\label{Intensity} The intensity and the diffraction of a square-integrable ergodic random measure $p$ are related by the formula
\[
\widehat{\eta}^+_p(\{1\}) = i(p)^2.
\]
\end{lemma}
For the proof of the intensity formula in the lemma above we consider a strongly continuous unitary $G$-representation $(\cH,\pi)$ and denote by $\Proj_G$ the projection onto the space of $G$-invariants. We then say that a sequence $(\beta_n)$ of probability measures on $G$ is \emph{weakly ergodic} if for every such representation $(\cH,\pi)$ we have convergence
\[
\lim_{n \ra \infty} \langle \pi(\beta_n)u,u \rangle = \|\Proj_G(u)\|^2_{\cH},  \quad \textrm{for all $u \in \cH$}.
\]
Weakly ergodic sequences of boundedly supported measures exist on every LCA group. For example, if $(F_n)$ is any F\o lner sequence in $G$, then by the weak mean ergodic theorem,  the sequence
\[
\dd\beta_n =\frac{\chi_{F_n}}{m_G(F_n)} \,  \dd m_G
\]
is weakly ergodic. In the case $G = \R^n$ we may e.g.\ choose $F_n$ to be the Euclidean ball of radius $n$ around $0$. If $(\beta_n)$ is any weakly ergodic sequence of probability measures on $G$ and if $1 \in \widehat{G}$ denotes the trivial character, then by definition we have
\begin{equation}\label{limM}
\lim_{n \ra \infty} \widehat{\beta}_n(\xi) = 0 \quad \text{ for all } \xi \in \widehat{G} \setminus\{1\}.
\end{equation}
\begin{proof}[Proof of Lemma \ref{Intensity}] Let $p: (\Omega, \bP) \to M(G)$ be a random measure; we consider the unitary representation $\pi$ of $G$ on $\cH := L^2(\Omega, \bP)$. Since $\bP$ is ergodic, for every $f \in C_c(G)$ the associated linear statistic $\cP f$ satisfies
\[
\Proj_G(\cP f) = \int_{\Omega} \cP f\, \dd \bP = \bE[p_\omega(f)] = M^1_p(f) = i_p \cdot m_G(f).
\]
Now let $(\beta_n)$ be a weakly ergodic sequence such that $\beta_n$ has bounded support for every $n$.  Then, for every non-negative $f \in C_c(G)$ and all $n \in \bN$  we have $\beta_n \ast f \ast f^* \in C_c(G)$ and,  by polarization,  
\begin{eqnarray*}
 \iota_p^2 \cdot  |\widehat{f}(1)|^2 &=& \iota_p^2 \cdot \left(\int_G f \,  dm_G \right)^2 \quad = \quad \|\Proj_G(\cP f)\|^2\quad=\quad  \lim_{n \ra \infty} \langle \pi(\beta_n) \cP{f}, \cP{f} \rangle_{L^2(\Omega, \bP)}\\
 &=&  \lim_{n \ra \infty} \eta_p^+(\check\beta_n\ast f \ast f^{*}) \quad = \quad  \lim_{n \ra \infty} \widehat{\eta}^+_p\left(\overline{\widehat{\beta}_n} \cdot |\widehat{f}|^2\right).
\end{eqnarray*}
Now ${\widehat{\beta}_n} \to \delta_1$ by \eqref{limM}, and since $\widehat{f} \in L^2(\widehat{\eta}^+_p)$ we can apply dominated convergence to obtain
\[
 \iota_p^2 \cdot  |\widehat{f}(1)|^2=  \widehat{\eta}^+_p(\{1\}) \cdot |\widehat{f}(1)|^2.
\]
If we choose $f$ with $\widehat{f}(1) \neq 0$, then we can cancel $|\widehat{f}(1)|^2$ and obtain $ \iota_p^2 =   \widehat{\eta}^+_p(\{1\})$.
\end{proof}
In the sequel, we will be interested mostly in the Euclidean case where $G = \R^d$ for some (arbitrary) dimension $d \in \mathbb N$. In this case we will identify $G$ with $\widehat{G}$ by identifying $\xi \in G$ with the character $x \mapsto e^{2\pi i \langle x, \xi \rangle}$. Under this identification we may then choose both $m_G$ and $m_{\widehat{G}}$ to be $d$-dimensional Lebesgue measure $\Vol_d$. We will later need the following a priori estimate concerning the diffraction of large balls.
\begin{lemma}[Dimension estimate]
\label{volgrowth}
Let $p$ be a square-integrable invariant random measure on $\R^d$ and denote by $B_R$ the Euclidean ball in $\R^d$. Then the centered diffraction $\widehat{\eta}_p$ satisfies the estimate $\widehat{\eta}_p(B_R) \ll_d R^d$ for all $R\geq 1$.  
\end{lemma}
\begin{proof}
For every $R\geq 1$,  we can find a finite subset $F_R \subset {\bR}^d$
such that 
\[
|F_R| \ll_d R^d \qand B_R \subset \bigcup_{\xi \in F_R} B(\xi,1) =  \bigcup_{\xi \in F_R} (B_1 + \xi).
\]
Hence, 
\[
\widehat{\eta}_p(B_R) \ll_d R^d \,  \sup_{\xi \in \bR^d} \widehat{\eta}_p(B_1 + \xi).
\]
Since $\eta_p$ is positive definite, \cite[Prop.\ 4.9]{BF} implies that $\widehat{\eta}_p$ is translation-bounded,  and thus the supremum on the right-hand side is bounded.  
\end{proof}

\subsection{Diffraction formulas}
To illustrate the above definitions, we recall two examples of explicit diffraction formulas, one for Poisson processes and one for cut-and-project processes. For the convenience of the reader we include proofs in Appendix \ref{DiffFormulas}. We first consider the case of a \emph{Poisson process}. Recall that if $(Y,m)$ is a $\sigma$-finite Borel measure space, then a Borel probability 
measure $\mu$ on the space of $\sigma$-finite Borel measures on $Y$ is called $m$-\emph{Poisson} if
\begin{itemize}
\item[(i)] for every Borel set $B \subset Y$ with finite and positive $m$-measure,  
\[
\mu\left(\left\{ p \in M_\sigma(Y) \,  : \,  p(B) = k \right\}\right) = 
\frac{m(B)^k \,  e^{-m(B)}}{k!},  \quad \textrm{for all $k \in \bN_o$}.
\]
\item[(ii)] for every $r \geq 1$,  and for all \emph{disjoint} Borel 
sets $B_1,\ldots,B_r \subset Y$ the corresponding linear statistics $\cP \chi_{B_1}$, \dots, $\cP \chi_{B_r}$
are $\mu$-independent.  
\end{itemize}
A point process is then called an \emph{$m$-Poisson process} if its distribution is $m$-Poisson. Such a process exists for every $\sigma$-finite Borel measure space $(Y,m)$ (see \cite[Theorem 3.6]{LP}) and is unique up to equivalence (see \cite[Prop.\ 3.2]{LP}). Here we will be interested in the case where $(Y, m) = (G, m_G)$. In this case it follows from invariance of $m_G$ that the $m_G$-Poisson measure is also $G$-invariant, hence there is an invariant $m_G$-Poisson process $p$, unique up to equivalence.
\begin{proposition}[Poisson diffraction]\label{DiffPoiss} For any lcsc group $G$ the following hold.
\begin{enumerate}[(i)]
\item The centered autocorrelation of the $m_G$-Poisson process $p$ on $G$ is given by $\eta_p = \delta_e$.
\item If $G$ is abelian, then the centered diffraction is given by  $\widehat{\eta}_p = m_{\widehat{G}}$.
\end{enumerate}
\end{proposition}
Here, the normalization of $m_{\widehat{G}}$ is determined by \eqref{DualMeasures}.
For example, if $(G, m_G) = (\R^d, \Vol_d)$, then the Poisson process satisfies
\begin{equation}\label{EuclideanPoisson}
\widehat{\eta}_p(B_r) = \Vol_d(B_r) = \Vol_d(B_1) \cdot r^d.
\end{equation}
In particular, the centered diffraction of any invariant Poisson process on $\R^d$ is absolutely continuous with respect to Lebesgue measure. On the contrary, we will see that the centered diffraction of any cut-and-project process is pure point. To make this precise we consider a cut-and-project process $p = p(G, H, \Gamma, W)$ as in Example \ref{ExCuP} with the additional assumption that $G$ and $H$ are abelian. We use the notation from Example \ref{ExCuP}, so that in particular $\Omega = \Gamma\backslash L$, where $L := G \times H$. We fix Haar measures $m_G$ and $m_H$ on $G$ and $H$ respectively and denote $m_L := m_G \otimes m_H$. We also fix a Borel fundamental domain $\cF \subset L$ for $\Gamma$; then $\mathrm{covol}(\Gamma) := m_L(\cF)$ depends only on $\Gamma$. Finally, we denote by 
\[
\widehat{L} = \widehat{G} \times \widehat{H} \qand \Gamma^\perp := \{ \xi \in \widehat{L} \,  : \,  \xi|_\Gamma = 1 \}
\]
the Pontryagin dual of $L$ and dual lattice of $\Gamma$ respectively. We then have the following formula, which, in essence, goes back to Meyer \cite{Meyer}.
\begin{theorem}[Cut-and-project diffraction]\label{DiffractionFormula}
The diffraction $\widehat{\eta}^+_p$ of $p = p(G, H, \Gamma, W)$ satisfies
\[
\widehat{\eta}^+_p =  \frac{1}{\mathrm{covol}(\Gamma)^2}  \cdot \sum_{\xi = (\xi_1, \xi_2) \in \Gamma^\perp} |\widehat{\chi}_W(\xi_2)|^2 \cdot \delta_{\xi_1},
\]
and consequently, the centered diffraction  $\widehat{\eta}_p$ is given by
\[
\widehat{\eta}_p =  \frac{1}{\mathrm{covol}(\Gamma)^2}  \cdot \sum_{\xi = (\xi_1, \xi_2) \in \Gamma^\perp \setminus \{(0,0)\}} |\widehat{\chi}_W(\xi_2)|^2 \cdot \delta_{\xi_1}.
\]
\end{theorem}

\section{Definitions of hyperuniformity}\label{SecHU}
\subsection{Spectral vs.\ geometric hyperuniformity}
Consider a locally square integrable invariant random measure $p: \Omega \to \mathrm{M}(\R^d)$ with associated centered autocorrelation $\eta_p$ and associated centered diffraction $\widehat{\eta}_p$. It follows from Proposition \ref{RemoveAtom}, that if $(U_n)$ is a nested sequence of compact identity neighbourhoods in $\widehat{G}$ with $\bigcap U_n = \{1\}$, then
\begin{equation}\label{PreHyper}
\lim_{n \to \infty} \widehat{\eta}_p(U_n) = 0,
\end{equation}
and (spectral) hyperuniformity is concerned with the speed of this convergence. Given $t>0$ and a subset $W \subset \widehat{\R}^d$ we write $tW := \{tx \mid x \in W\}$.

\begin{definition} Let $W \subset \widehat{\R}^d$ be a bounded identity neighbourhood. The random measure $p$ is \emph{spectrally hyperuniform with respect to $W$} if
\[
\lim_{t \to 0} \frac{\widehat{\eta}_p(tW)}{t^d} = 0.
\]
It is called \emph{spectrally hyperuniform} if it is spectrally uniform with respect to the Euclidean unit ball in $\widehat{\R}^d$.
\end{definition}
In view of \eqref{EuclideanPoisson}, the denominator $t^d$ can be interpreted (up to a constant) either as the Lebesgue volume of $tW$ or as the diffraction measure of $tW$ with respect to a Poisson process. Hyperuniformity thus corresponds to ``sub-Poissonian'' spectral behaviour near $0$. A dual approach to hyperuniformity, which is often crucial for applications, is via the \emph{number variance} of $p$. 
\begin{definition}  Let $V \subset {\R}^d$ be a bounded identity neighbourhood. The random measure $p$ is \emph{geometrically hyperuniform with respect to $V$} if
\[
\lim_{t \to \infty} \frac{\mathrm{Var}_p(tV)}{t^d} = 0.
\]
It is called \emph{geometrically hyperuniform} if it is geometrically uniform with respect to the Euclidean unit ball in $\R^d$.
\end{definition}
\begin{proposition}[Spectral vs.\ geometric hyperuniformity]\label{SvsG} A locally square integrable invariant random measure is spectrally hyperuniform if and only if it is geometrically hyperuniform.
\end{proposition}
Special cases of Proposition \ref{SvsG} have been obverved in different levels of generality by many people, see e.g.\ \cite[Prop.\ 2.2]{C}. Lacking a reference in the present generality, we will include a full proof. In fact, we will provide a more precise version in Theorem \ref{TheoremSvsG} below; see also Theorem \ref{Quanti} for a quantitative version.

.  It is important to note that a hyperuniform invariant random measure need not be geometrically hyperuniform with respect to balls of a \emph{non-Euclidean} metric, as the following example shows.
\begin{example} 
\label{ExampleDim2} Let $\Gamma < \R^d$ be a lattice and let $\delta: \R^d/\Gamma \to M(\R^d)$ be the associated periodic simple point process.
We claim that $\delta$ is spectrally hyperuniform. Indeed, by Poisson summation the support of its centered diffraction is $\Gamma^\perp \setminus \{0\}$, where $\Gamma^\perp$ denotes the dual lattice of $\Gamma$, and thus the centered diffraction vanishes in a neighbourhood of $0$. By Proposition \ref{SvsG} it is thus 
geometrically hyperuniform with respect to Euclidean balls. On the other hand, even for $\Z^2 \subset \R^2$ this process is not geometrically hyperuniform with respect to $\ell^\infty$-balls in $\R^2$, see \cite[Section 2.1]{C}. However, $\ell^\infty$-balls are not Fourier smooth in the sense of the following remark.
\end{example}
\begin{remark}[Fourier smoothness of balls]\label{DecayWindows} 
If $B_r$ denotes the Euclidean ball in $\R^d$, then its Fourier transform satisfies the estimate
\[
|\widehat{\chi}_{B_r}(\xi)| \ll (1 + \|\xi\|)^{-(d+1)/2}  \quad \textrm{for all $\xi \in \widehat{\R}^d$},
\]
In this article we will often consider the wider class of Borel sets $B \subset \bR^d$ whose Fourier transforms satisfy the estimate
\[
|\widehat{\chi}_B(\xi)| \ll (1 + \|\xi\|)^{-(d+\vartheta)/2}  \quad \textrm{for some $\vartheta > 0$ and all $\xi \in \widehat{\R}^d$}.
\]
Such sets will be called \emph{Fourier smooth} with \emph{exponent} $\vartheta$ in the sequel. With this terminology, Euclidean balls are thus Fourier smooth with exponent $1$. By \cite[Thm.\ 2.16]{I}, a compact, convex and symmetric subset $B \subset \R^d$ is Fourier smooth if its boundary is $(d+3)/2$-times differentiable and its principal curvatures do not vanish. On the other hand, $\ell^\infty$-balls in $\R^2$ are not Fourier smooth.\end{remark}
As the following theorem shows, the problems encountered above with $\ell^\infty$-balls do not occur for Fourier smooth sets.
\begin{theorem}\label{TheoremSvsG}
Let  $p: \Omega \to \mathrm{M}(\R^d)$ be a locally square integrable invariant random measure.
\begin{enumerate}[(i)]
\item If $p$ is geometrically hyperuniform with respect to some bounded Borel set $V \subset \R^d$, then it is spectrally hyperuniform.
\item If $p$ is spectrally hyperuniform for some bounded Borel set $W \subset \widehat{\R}^d$ with $0$ in its interior, then it is spectrally hyperuniform with respect to any such set and in particular spectrally hyperuniform.
\item If $p$ is spectrally hyperuniform, then it is geometrically hyperuniform with respect to every Fourier smooth bounded Borel set $V$.
\end{enumerate}
\end{theorem}
\begin{proof} We are going to use the fact that, by \eqref{Diffraction} and since $\widehat{\chi}_{tV}(\xi) = t^d \cdot \widehat{\chi}_V(t\xi)$, we have
\begin{equation}\label{VartV}
\Var_p(tV) = \widehat{\eta}_p(|\widehat{\chi}_{tV}|^2) = t^{2d} \cdot  \int_{\bR^d} |\widehat{\chi}_V(t\xi)|^2 \,  \dd\widehat{\eta}_p(\xi) 
\end{equation}
for any bounded Borel set $V \subset \R^d$ and every $t>0$.

\item (i) Since $V$ is bounded, the Fourier transform $\widehat{\chi}_V$ is continuous. Since moreover $\widehat{\chi}_V(0) = \Vol_d(V)$, there thus exists a constant $c > 0$ such that
\[
|\widehat{\chi}_V(\xi)|^2 \geq \frac{\Vol_d(V)^2}{2} \quad \textrm{for all $\xi \in B_c$}.
\]
Using \eqref{VartV} and the fact that $\widehat{\eta}_p$ is a positive measure we have for all $t \geq 1$,
\[
\frac{\Var_p(tV)}{t^d} \;=\; t^d \cdot \int_{\bR^d} |\widehat{\chi}_{V}(t\xi)|^2 \,  \dd\widehat{\eta}_p(\xi)\;\geq\; t^d \cdot \int_{B_{c/t}}  |\widehat{\chi}_V(t\xi)|^2 \,  \dd\widehat{\eta}_p(\xi) \;\geq\; \frac{\Vol_d(V)^2}{2} \cdot t^d \cdot \widehat{\eta}_p(B_{c/t}),
\]
i.e.\ for $t \geq 1$ we have
\[
\widehat{\eta}_p(B_{c/t}) \leq \frac{2}{t^{2d}} \cdot \frac{\Var_p(tV)}{\Vol_d(V)^2}.
\]
Setting $\varepsilon := c/t$ this yields
\[
\lim_{\eps \ra 0^{+}} \frac{\widehat{\eta}_p(B_\eps(0))}{\eps^d} \leq \lim_{t \to \infty}  \frac{t^d}{c^d}\cdot\frac{2}{t^{2d}} \cdot \frac{\Var_p(tV)}{\Vol_d(V)^2} = \frac{2}{c^d \cdot \Vol_d(V)^2} \cdot \lim_{t \to \infty} \frac{\Var_p(tV)}{t^d}  = 0.
\]
\item (ii) If $W$ and $W'$ are bounded Borel sets with $0$ in its interior, then there exist $R>r>0$ such that
\[
rW \subset W' \subset RW \implies r\cdot \frac{\widehat{\eta}_p(rtW)}{rt}\leq \frac{\widehat{\eta}_p(tW')}{t}\leq R \cdot\frac{\widehat{\eta}_p(RtW)}{Rt} \text{ for all }t>0.
\]
This shows that
\[
\lim_{t \to \infty}  \frac{\widehat{\eta}_p(tW)}{t} = 0 \iff \lim_{t \to \infty}  \frac{\widehat{\eta}_p(tW')}{t} = 0.
\]
\item (iii) In view of \eqref{VartV} and the assumption of Fourier smoothness of $V$ we have for all $R >1$,
 \begin{align*}
\frac{\Var_p({RV})}{R^d}
&= 
R^d \cdot  \int_{\bR^d} |\widehat{\chi}_B(R\xi)|^2 \,  \dd\widehat{\eta}_p(\xi) \\[0.2cm]
&\ll 
R^d \cdot \int_{\bR^d} (1 + R\|\xi\|)^{-(d+\vartheta)} \,  \dd\widehat{\eta}_p(\xi) \\[0.2cm]
&=
R^d \cdot \int_0^1 \widehat{\eta}_p
\left( \left\{ \xi \in \bR^d \,  : \,  (1 + R\|\xi\|)^{-(d+\vartheta)} \geq t \right\}\right) \,  \dd t \\[0.2cm]
\end{align*}
If we set $\Psi(s) := (1+s)^{-(d+\vartheta)}$ and $u := R^{-1} \Psi^{-1}(t)$, then the condition under the integral is given by
\[
\Psi(R\|\xi\|) \geq t \iff \|\xi\| \leq u \iff \xi \in B_u, \quad \text{and} \quad \dd t = R \cdot \Psi'(Ru)\, du = \frac{(-d-\vartheta) R}{(1+Ru)^{d+\vartheta + 1}}\, \dd u,
\]
and hence the substitution $t \mapsto u$ yields
\begin{equation}\label{IntegralTriEstimate}
\frac{\Var_p(RV)}{R^d} \ll R^{d+1} \cdot \int_0^\infty F_{R,d, \vartheta}(u) \,\dd u, \quad \text{where} \quad F_{R,d, \vartheta}(u) := \frac{\widehat{\eta}_p\left(B_u\right)}{(1 + Ru)^{d + \vartheta + 1}}.\end{equation}
To estimate the integral on the right, we will break the domain of integration into three parts. From now on we fix $\eps > 0$. We then choose $q$ in the open interval $(0, \frac{\vartheta}{d+\vartheta+1})$. Since $q > 0$  we can find, by spectral hyperuniformity, a constant $R_\eps(q)$ such that
\[
\widehat{\eta}_p(B_{t/R}) \leq \eps \left(t/R \right)^d \quad \text{for all $R \geq R_\eps(q)$ and for all $0 \leq t \leq R^{1-q}$}.\] 
Using Lemma \ref{volgrowth} we have
\[
\widehat{\eta}_p(B_u) \ll u^{d} \quad \textrm{for all $u \geq M \geq 1$}.
\]
For all $R \geq R_\eps(q)$ we then have, by our definition of $R_\eps$,
\begin{equation}\label{IntegralTriEstimate1}
I_1 :=  R^{d+1} \cdot \int_0^{R^{-q}} F_{R,d, \vartheta}(u) \, \dd u =  R^d \int_0^{R^{1-q}} \frac{\widehat{\eta}_p(B_{t/R})}{(1 + t)^{d+ \vartheta+1}} \,  \dd t  \leq \eps  \int_0^{R^{1-q}} \frac{t^d}{(1 + t)^{d+\vartheta+1}} \,  \dd t \ll \eps.
\end{equation}
for all $R \geq R_\eps(q)$. Secondly, by our choice of $q$ we have
\begin{equation}\label{IntegralTriEstimate2}
 I_2 :=  R^{d+1} \cdot \int_{R^{-q}}^M F_{R,d, \vartheta}(u) \, \dd u \leq \frac{R^{d+1} \cdot M \cdot \widehat{\eta}_p(B_M)}{(1+R^{1-q})^{d+\vartheta+1}} \xrightarrow{R \to \infty} 0,
\end{equation}
and, finally, by our choice of $M$ we have 
\begin{equation}\label{IntegralTriEstimate3}
I_3 :=  R^{d+1} \cdot \int_M^\infty F_{R,d, \vartheta}(u) \, \\d u \ll 
R^{d+1} \int_{M}^\infty \frac{t^d}{(1+Rt)^{d+\vartheta+1}} \,  \dd t \; =\;
\int_{RM}^\infty \frac{t^d}{(1+t)^{d+\vartheta+1}} \,  \dd t \xrightarrow{R \to \infty} 0.
\end{equation}
Plugging \eqref{IntegralTriEstimate1}, \eqref{IntegralTriEstimate2} and \eqref{IntegralTriEstimate3} into \eqref{IntegralTriEstimate} then yields
\[
\varlimsup_{R \to \infty} \frac{\Var_p(RV)}{R^d} \ll \eps,
\]
and since $\eps>0$ was chosen arbitrarily, the theorem follows.
\end{proof}
Note that (i) and (iii) imply Proposition \ref{SvsG} (and hence Theorem \ref{EquivHU} from the introduction). In the sequel we say that $p$ is \emph{hyperuniform} if it is spectrally, or equivalently geometrically hyperuniform. 

\subsection{Quantitative bounds}

Let $V$ be a bounded Borel set containing $0$ in its interior. If $V$ is Fourier smooth, then Theorem \ref{TheoremSvsG} states that
\[
\lim_{R \to \infty} \frac{\Var_p(RV)}{\Vol_d(RV)} = 0 \iff \lim_{\eps \to 0} \frac{\widehat{\eta}_p(B_\eps)}{\Vol_d(B_\eps)} = 0.
\]
From the proof one can actually obtain a more quantitative relation between the variance of large balls and the centered diffraction measure of small balls.  We collect
these relations in the following theorem,  and leave the proof to the reader. 

\begin{theorem}\label{Quanti} Let $p$ be a locally square-integrable random measure and let $V$ be a bounded Borel set containing $0$ in its interior.
\begin{enumerate}[(i)]
\item There is a constant $c_V > 0$ such that for every function $\rho : [1,\infty) \ra (0,\infty)$,
\[
\Var_p(RV) = O(R^d \rho(R)),  \enskip R \ra \infty \implies \widehat{\eta}_p(B_\eps) = O(\eps^d \rho(c_V/\eps)),  \enskip \eps \ra 0^{+}.
\]
\item If $V$ is Fourier smooth with exponent $\vartheta$ and $0 \leq \gamma < \vartheta$, then 
\[
\Var_p(RV) = O(R^{d-\gamma}),  \enskip R \ra \infty
\Longleftrightarrow 
\widehat{\eta}_p(B_\eps) = O(\eps^{d+\gamma}),  \enskip \eps \ra 0^{+}.
\]
\end{enumerate}
The same statements hold for $o$ instead of $O$.
\end{theorem}

\subsection{Counterexamples in dimension $1$}
Theorem \ref{TheoremSvsG}.(iii) rests on Fourier smoothness of the set $V$. Example \ref{ExampleDim2} shows that this assumption is indeed necessary in dimensions $d \geq 2$. To see that it is also necessary in dimension $d = 1$ one can use a construction of Brown, Glicksberg and Hewitt \cite{BGH}. More precisely,  we show that for a large class of one-dimensional locally square integrable invariant random measures (including periodic and quasi-crystalline ones),  there is always a compact subset of the real line with respect to which these processes are \emph{not} geometrically hyperuniform.  The following result can be found in \cite[Example C]{BGH}.
\begin{lemma}
\label{Lemma_BGH}
There exist a compact subset $V \subset \bR$ and a sequence $(\xi_n)$ in $(0,\infty)$ such that
$\xi_n \ra +\infty$ and
\[
 \lim_{n \ra \infty} \xi_n^{1/2} \cdot |\widehat{\chi}_V(\xi_n)| > 0,  \quad \textrm{as $n \ra \infty$}.
\]
\end{lemma}

Note that $V$ is \emph{not} Fourier smooth for any positive exponent.  We can now prove:

\begin{lemma}[Automatic geometric non-hyperuniformity]
Let $V$ and $(\xi_n)$ be as in Lemma \ref{Lemma_BGH},  and fix a point $\xi_o \in (0,\infty)$.  Then,  for every locally square integrable invariant random measure $p : \Omega \ra M(\bR)$
for which $\xi_o$ is an atom of $\widehat{\eta}_p$,  we have
\[
\limsup_{n \ra \infty} \frac{\Var_p(R_n V)}{R_n} > 0,
\]
where $R_n = \frac{\xi_n}{\xi_o} \ra \infty$ as $n \ra \infty$.
\end{lemma}

\begin{proof}
Note that if $R_n = \xi_n/\xi_o$,  then
\begin{align*}
\frac{\Var_p(R_n V)}{R_n} 
&= R_n \cdot \int_{-\infty}^\infty |\widehat{\chi}_V(R_n \xi)|^2 \,  d\widehat{\eta}_p(\xi) \geq R_n \cdot |\widehat{\chi}_V(R_n \xi_o)|^2 \cdot \widehat{\eta}_p(\{\xi_o\}) \\[0.2cm]
&= \frac{\widehat{\eta}_p(\{\xi_o\})}{\xi_o} \cdot \left( \xi_n^{1/2} \cdot |\widehat{\chi}_V(\xi_n)| \right)^2.
\end{align*}
By our assumptions on $V$ and $(\xi_n)$,  the limsup of the right-hand side is strictly positive,  and the proof is done.
\end{proof}

\section{Cut-and-project processes which are not hyperuniform}\label{Counter}
\subsection{A class of cut-and-project processes}
Given dimension parameters $d_1$, $d_2$, every choice of lattice $\Gamma<\R^{d_1+d_2}$ and window $W \subset \R^{d_2}$ gives rise to a cut-and-project process  $p(\R^{d_1}, \R^{d_2}, \Gamma, W)$ (see Example \ref{ExCuP}).  In this section we are going to show that,  already in the smallest possible case where $d_1 = d_2 = 1$, it is possible to choose the lattice $\Gamma$ and the window $W$ in such a way,  that the resulting point process is non-hyperuniform in a very strong quantitative sense. We are going to choose lattices of the form \[\Gamma_a := g_a\bZ^2 < \R^2, \quad \text{where}\quad
g_a 
= 
\frac{1}{2a}
\left(
\begin{matrix}
1 & - a \\
1 & a
\end{matrix}
\right) \quad \text{for some} \quad a > 0,
\]
and windows for the form $W_b := [-b,b]$ for some $b>0$. We will choose $a \in \R \setminus \Q$, so that $\Gamma_a$ is irreducible.
We then denote by $p_{a,b} = p(\bR,\bR,\Gamma_a, W_b)$ the corresponding cut-and-project process. As a special case of  Theorem \ref{DiffractionFormula} we have:
\begin{corollary}\label{Diffab} Let $a,b > 0$ with $a \in \bR \setminus \Q$. Then the centered diffraction $\widehat{\eta}_{a,b}$ of the process $p_{a,b}$ is given by
\begin{equation}
\label{CPdiffGammaab}
\widehat{\eta}_{a,b}([-u,u])
=
4a^2 \cdot \sum_{(m,n) \in \bZ^2 \setminus \{(0,0\}} \chi_{[-1,1]}\left(\frac{am-n}{u}\right) \cdot \left|\widehat{\chi}_{[-b,b]}(am+n)\right|^2.
\end{equation}
\end{corollary}
\begin{proof} Since $\det(g_a) = \frac{1}{2a}$ and $\bZ^2$ is unimodular, we have $\covol(\Gamma_a) = \frac{1}{2a}$, hence it suffices to observe that
\[
g_a^{-T} = 
\left(
\begin{matrix}
a & - 1 \\
a & 1
\end{matrix}
\right)
\qand
\Gamma_a^{\perp}
= g_a^{-T} \bZ^2 = 
\left\{ 
\left(
\begin{matrix}
am - n \\
am + n
\end{matrix}
\right)
\,  : \,  m, n \in \bZ
\right\}.
\qedhere\]
\end{proof}
\subsection{Choice of parameters}
It turns out that the question whether the process $p_{a,b}$ defined above is hyperuniform or not depends on diophantine properties of the parameters $a$ and $b$. Given a real number $\theta$, we set
\[
\{\theta\}_{\bZ} = \min\{ |\theta + n| \,  : \,  n \in \bZ \} \in [0,1/2].
\]
We will need $a$ to be irrational, but well-approximable by rational numbers. More precisely, we are going to assume that $a \in \R \setminus \Q$ and that there exists some $\gamma > 2$ and an integer sequence $(m_k)$ such that
\begin{equation}
\label{ass_mk}
m_k \ra \infty \qand \{m_k a\}_{\bZ} \leq m_k^{-\gamma} \quad \textrm{for all $k$}.
\end{equation}
We recall that $a$ is called a \emph{Liouville number} if such a sequence $(m_k)$ exists in fact for \emph{every} $\gamma > 2$; it is well-known that there exist uncountably many Liouville numbers.

We are going to show that if $a \in \R \setminus \Q$ satisfies \eqref{ass_mk}, then for Lebesgue almost every $b \in (0,1/2a]$,  we have
\begin{equation}
\label{nonhypCPlimsup}
\varlimsup_{k \ra \infty} \frac{\widehat{\eta}_{a,b}([-u,u]}{u^{\delta}} = \infty,  
\quad \textrm{for all $\delta > \frac{2}{\gamma}$}.
\end{equation}
In particular, if $a$ is a Liouville number, then \eqref{nonhypCPlimsup} holds for all $\delta > 0$, and thus we have established Theorem \ref{AntiHUIntro} from the introduction. In fact, we have the following
slightly more precise version of \eqref{nonhypCPlimsup}.
\begin{theorem}
\label{Thm_NonHypCP}
Let $\gamma > 2$ and suppose that $a$ satisfies \eqref{ass_mk} for some sequence $(m_k)$.  Let $u_k = 2 m_k^{-\gamma}$.  Then there is a Lebesgue-conull subset $E_a \subset (0,1/2a]$ such that for every 
$b \in E_a$,  there is a sub-sequence $(k_j)$ with the property that for all 
$\delta > \frac{2}{\gamma}$, 
\[
\lim_{j \ra \infty} \frac{\widehat{\eta}_{a,b}([-u_{k_j},u_{k_j}])}{u_{k_j}^{\delta}} = \infty.
\]
\end{theorem}

\subsection{Proof of Theorem \ref{Thm_NonHypCP}}

Theorem \ref{Thm_NonHypCP} is a consequence of the following two lemmas.
\begin{lemma}
\label{Lemma_LowerBndEu}
For $u \in (0,1)$,  define 
\[
Q_u = \left\{ m \in \bZ \setminus \{0\} \,  : \,  \{am\}_{\bZ} \leq u/2 \right\}.
\]
Then,  for all $a \in \bR \setminus \bQ$,   $b \in \bR$ and $0 < u <  1$,  
\[
\widehat{\eta}_{a,b}([-u,u]) 
\geq \frac{1}{2} \cdot \sum_{m \in Q_u} \left( \frac{\sin(4\pi abm)}{\pi m} \right)^2  +O_{a,b}(u).
\]
\end{lemma}

The next lemma explains how the sequence $(k_j)$ is chosen. 

\begin{lemma}
\label{Lemma_Equi}
For every $a \in \bR \setminus \{0\}$ and sequence $(m_k)$ of integers such that $m_k \ra \infty$ as $k \ra \infty$,  there exist a Lebesgue conull subset $E_a \subset (0,1/2b]$ with the property that for every $b \in E_a$,  there is a sub-sequence $(k_j)$ such that
\[
\lim_{j \ra \infty} \sin(4\pi ab m_{k_j}) = 1.
\]
\end{lemma}

\begin{proof}[Proof of Theorem \ref{Thm_NonHypCP} assuming Lemma \ref{Lemma_LowerBndEu} and Lemma \ref{Lemma_Equi}]
Let $\gamma > 2$ and $(m_k)$ be as in Theorem \ref{Thm_NonHypCP},  and let $u_k := 2m_k^{-\gamma}$.  Note that $u_k \ra 0$ as $k \ra \infty$ and $m_k \in Q_{u_k}$ for all $k$ and thus,  
by Lemma \ref{Lemma_LowerBndEu},
\[
\widehat{\eta}_{a,b}([-u_k,u_k]) \geq  \frac{1}{2} \cdot \left( \frac{\sin(4\pi abm_k)}{\pi m_k} \right)^2 + 
O_{a,b}\left(m_k^{-\gamma}\right),  \quad \textrm{for  all $k$}.
\]
In particular,  for every $\delta \in (0,1)$,
\[
\frac{\widehat{\eta}_{a,b}([-u_k,u_k])}{u_k^\delta}
\geq 2^{-(1+\delta)} \cdot m_k^{\delta\gamma-2} \cdot \sin^2(4\pi ab m_k) + 
O_{a,b}\left(m_k^{-(1-\delta)\gamma}\right),
\]
for all $k$.  Note that since $0 < \delta < 1$,  the $O$-term tends to zero when $k \ra \infty$. \\

Let $E_a$ be as in Lemma \ref{Lemma_Equi},  and fix $b \in E_a$.  The same lemma allows us to extract a sub-sequence $(k_j)$ such 
that $\lim_{j \ra \infty} \sin^2(4\pi ab m_{k_j}) = 1$.  We note that if $\delta > \frac{2}{\gamma}$,  then $m_{k_j}^{\delta\gamma-2} \ra \infty$
when $j \ra \infty$,  and thus
\[
\lim_{j \ra \infty}\frac{\widehat{\eta}_{a,b}([-u_{k_j},u_{k_j}])}{u_{k_j}^\delta}
\geq 2^{-(1+\delta)} \cdot \lim_{j \ra \infty} m_{k_j}^{\delta\gamma-2} \sin^2(4\pi ab m_{k_j}) = \infty,
\]
which finishes the proof.
\end{proof}
Lemma \ref{Lemma_Equi} follows from a standard equidistribution argument:
\begin{proof}[Proof of Lemmma \ref{Lemma_Equi}]
Let $(m_k)$ be an integer sequence such that $m_k \ra \infty$ when $k \ra \infty$.  
By \cite[Chapter 1,  Theorem 4.1]{KN} we can find a conull subset $E \subset (0,1)$ such that  $(m_{k}\alpha)$ is equidistributed modulo $1$ for every $\alpha \in E$. In particular, for every $\alpha \in E$,  we can find a subsequence $(k_j)$ such that $m_{k_j} \alpha \ra \frac{1}{4} \mod 1$. Define
\[
E_a := \left\{b \in \left(0,\frac 1{2a}\right) \mid 2ba \in E\right\}
\]
Then for every $b \in E_a$ we find a sequence $(m_{k_j})$ such that $2bam_{k_j} \to \frac 1 4 \mod 1$ and hence  
\[\sin(4\pi ab m_{k_j}) \to \sin\left(2\pi \cdot \frac{1}{4}\right) = 1.\]
Since $E$ is conull in $(0,1)$,  the set $E_a$ is conull in $(0,1/2a)$,  and we are done.
\end{proof}
We have thus reduced the proof of the theorem to Lemma \ref{Lemma_LowerBndEu}. The remainder of this section is devoted to deducing this lemma from the diffraction formula (Corollary \ref{Diffab}).

\subsection{Proof of Lemma \ref{Lemma_LowerBndEu}}
\label{Subsec:PrfLowerBndEu}

For the proof of Lemma \ref{Lemma_LowerBndEu} it will be convenient to abbreviate
\begin{equation}
\label{Def_phi1phi2}
\varphi_1(x) := \chi_{[-1/2,1/2]} * \chi_{[-1/2,1/2]}(x) \qand \varphi_2(x) = |\widehat{\chi}_{[-b,b]}(x)|^2 := \left( \frac{\sin(2\pi b x)}{\pi x} \right)^2
\end{equation}
Note that
\[
\widehat{\varphi}_1(\xi) = \left( \frac{\sin(\pi \xi)}{\pi \xi} \right)^2
\qand
\widehat{\varphi}_2(\xi) = (\chi_{[-b,b]} * \chi_{[-b,b]})(\xi).
\]
From the diffraction formula we can derive a lower bounded for $\widehat{\eta}_{a,b}([-u,u])$ in terms of the functions
\begin{equation}
\label{Def_Hmu}
H_m(u) := u \cdot 
\sum_{k=-\infty}^{\infty} e^{-2\pi i am k} \cdot \int_{-2b}^{2b} \widehat{\varphi}_1(u(x+k)) \,  \widehat{\varphi}_2(x) e^{-4\pi i am x} \,  \dd x,  \quad \textrm{for $m \neq 0$}.
\end{equation}
\begin{lemma}
\label{Lemma_Lowerbndeta1}
For all $u > 0$,  
\[
\widehat{\eta}_{a,b}([-u,u]) \geq 4a^2 \cdot \sum_{m \neq 0} H_m(u).
\]
\end{lemma}
\begin{proof}
Since the diffraction formula \eqref{CPdiffGammaab} only involves non-negative terms,  we have
\begin{align*}
\widehat{\eta}_{a,b}([-u,u])
&=
4a^2 \cdot \sum_{(m,n) \in \bZ^2 \setminus \{(0,0\}} \chi_{[-1,1]}\left(\frac{am-n}{u}\right) \cdot \left|\widehat{\chi}_{[-b,b]}(am+n)\right|^2 \\[0.2cm]
&\geq 
4a^2 \cdot \sum_{m \neq 0} \left( \sum_{n=-\infty}^\infty \chi_{[-1,1]}\left(\frac{am-n}{u}\right) \cdot \left|\widehat{\chi}_{[-b,b]}(am+n)\right|^2 \right).
\end{align*}
We would like to apply Poisson's summation formula to the inner sum,  but since 
$\chi_{[-1,1]}$ does not have sufficient Fourier decay,  we cannot do this directly.
Instead,  using the simple estimate  $\chi_{[-1,1]} \geq \chi_{[-1/2,1/2]} * \chi_{[-1/2,1/2]}$,  we get a new lower bound: 
\[
\widehat{\eta}_{a,b}([-u,u]) \geq 4a^2 \cdot 
\sum_{m \neq 0} \left( \sum_{n=-\infty}^\infty \chi_{[-1/2,1/2]} * \chi_{[-1/2,1/2]}\left(\frac{am-n}{u}\right) \cdot \left|\widehat{\chi}_{[-b,b]}(am+n)\right|^2 \right).
\]
The functions in the inner sum on the right hand side do now have the right Fourier decay for for Poisson's summation formula to be applied.  For a fixed $m \neq 0$,  we denote the inner sum by $H_m(u)$,  and note that
\[
H_m(u) = \sum_{n=-\infty}^\infty \psi_{am}(n;u) 
\quad \textrm{where} \quad  \psi_\alpha(x;u) = \varphi_1\left(\frac{\alpha-x}{u}\right)\varphi_2(\alpha+x),  \enskip \alpha > 0.
\]
We leave it to the reader to check that
\[
\widehat{\psi}_\alpha(\xi;u) = u \cdot e^{-2\pi i \alpha \xi} \cdot \int_{-\infty}^{\infty} 
\widehat{\varphi}_1(u(x+\xi)) \cdot \widehat{\psi}_2(x) e^{-4\pi i \alpha x} \,  \dd x,
\]
and thus $H_m(u)$ is really given by \eqref{Def_Hmu}.
\end{proof}
The key estimate concerning the functions $H_m$ is as follows:
\begin{lemma}
\label{Lemma_Lowerbndeta2}
For all $u \in (0,1)$ and $m \neq 0$,  
\begin{align*}
H_m(u) 
&= 
\left(\sum_{n=-\infty}^\infty \varphi_1\left(\frac{am + n}{u}\right)\right) \cdot \varphi_2(2am) \\[0.2cm]
&-  \left(\sum_{n=-\infty}^\infty (am+n) \cdot \varphi_1\left(\frac{am + n}{u}\right)\right) \cdot \varphi_2'(2am) +O\left(\frac{u}{m^2}\right).
\end{align*}
\end{lemma}
Assuming this estimate for the moment, let us complete the proof:
\begin{proof}[Proof of Lemma \ref{Lemma_LowerBndEu} assuming Lemma \ref{Lemma_Lowerbndeta2}]
Upon combining these two lemmas,  we get
\begin{align}
\widehat{\eta}_{a,b}([-u,u])
&\geq 
4a^2 \cdot \sum_{m \neq 0}\left(\sum_{n=-\infty}^\infty \varphi_1\left(\frac{am + n}{u}\right)\right) \cdot \varphi_2(2am) \nonumber \\[0.2cm]
&-  4a^2 \cdot \sum_{m \neq 0}\left(\sum_{n=-\infty}^\infty (am+n) \cdot \varphi_1\left(\frac{am + n}{u}\right)\right) \cdot \varphi_2'(2am) +O\left(u\right).
\label{mergetwo}
\end{align}
For the first sum,  we note that $\varphi_1 \geq \frac{1}{2} \cdot \chi_{[-1/2,1/2]}$,  and thus the inner sum over $n$ is bounded from below by $\frac{1}{2} \cdot \chi_{Q_u}$.  Hence,
\begin{align*}
4a^2 \cdot \sum_{m \neq 0}\left(\sum_{n=-\infty}^\infty \varphi_1\left(\frac{am + n}{u}\right)\right) \cdot \varphi_2(2am) 
&\geq \frac{4a^2}{2} \cdot \sum_{m \in Q_u} 
\left( \frac{\sin(4\pi ab m)}{2\pi am} \right)^2 \\[0.2cm]
&= \frac{1}{2} \sum_{m \in Q_u} \left( \frac{\sin(4\pi ab m)}{\pi m} \right)^2.
\end{align*}
For the second sum,  we note that since $\supp(\varphi_1) \subset [-1,1]$ and
$\|\varphi_1\|_\infty \leq 1$,  the only terms which contribute to the inner sum 
are the ones for which $|am+n| \leq u$.  Since $u \in (0,1)$,  there are at most two such 
indices $n$.  Hence, 
\[
\left| 
\left(\sum_{n=-\infty}^\infty (am+n) \cdot \varphi_1\left(\frac{am + n}{u}\right)\right)
\right| \leq \sum_{\substack{n \in \bZ \\ |am+n| \leq u}} |am+n| \leq 2u,
\]
for all $m$.  Since $|\varphi_2'(x)| \ll x^{-2}$,  we see that $|\varphi_2'(2am)| \ll m^{-2}$,
and thus
\[
4a^2 \cdot \sum_{m \neq 0}\left(\sum_{n=-\infty}^\infty (am+n) \cdot \varphi_1\left(\frac{am + n}{u}\right)\right) \cdot \varphi_2'(2am) = O(u).
\]
Upon plugging this into \eqref{mergetwo},  and merging the two $O(u)$-terms,  we are done.
\end{proof}
We have thus reduced the proof of the theorem further to Lemma \ref{Lemma_Lowerbndeta2}, which is a purely analytic statement about the functions $\varphi_1, \varphi_2$ and their Fourier transforms.

\subsection{Proof of Lemma \ref{Lemma_Lowerbndeta2}}
We break the proof of Lemma \ref{Lemma_Lowerbndeta2} into four lemmas involving the auxiliary function
\[
F_\theta(x;u) = \sum_{k=-\infty}^\infty e^{-2\pi i \theta k} \left(\widehat{\varphi}_1(u(x+k)) - \widehat{\varphi}_1(uk) \right),  \quad x \in [-2b,2b],
\]
where $\theta \in \bR$ and $u > 0$ are parameters. 

\begin{lemma}
\label{Lemma_Hmu}
For all $u > 0$ and $m \neq 0$, 
\[
H_m(u) = 
\left(\sum_{n=-\infty}^\infty \varphi_1\left(\frac{am + n}{u}\right)\right) \cdot \varphi_2(2am)
+ u \cdot \int_{-2b}^{2b} F_{am}(x;u) \,  \widehat{\varphi}_2(x) \,  e^{-4\pi i am x} \,  \dd x
\]
\end{lemma}

\begin{lemma}
\label{Lemma_Ftheta}
For every $\theta \in \bR$ and $u > 0$,  
\begin{align*}
F_{\theta}(x;u) &= x \cdot \frac{2\pi i}{u} \cdot \sum_{n=-\infty}^\infty (n+\theta) \cdot \varphi_1\left(\frac{\theta + n}{u}\right) + \widehat{\varphi}_1(ux) - \widehat{\varphi}_1(0) \\[0.2cm]
&+ u^2 \cdot L_1(x;u,\theta)  - i u^3 \cdot L_2(x;u,\theta),
\end{align*}
for all $x \in [-2b,2b]$, where $L_1$ and $L_2$ are of the form
\[
L_1(x;u,\theta) = \int_0^x \int_{-y}^y G_1(z;u,\theta) \,  \dd z \,  \dd y
\]
and
\[
L_2(x;u,\theta) = \int_0^x \int_0^y \int_{-z}^z G_2(w;u,\theta) \,  \dd w \,  \dd z \,  \dd y,
\]
for certain  continuous functions $G_1,  G_2 : \bR \times \bR^{+} \times \bR \ra \bC$ which satisfy
\[
\sup\left\{ |G_j(x;u,\theta)| \,  : \,  x \in [-2b,2b],  \enskip u > 0,  \enskip \theta \in \bR \right\} \ll \frac{1}{u},  \quad j = 1,2.
\]
\end{lemma}

\begin{lemma}
\label{Lemma_VdC1}
For all $m \neq 0$,
\[
\sup_{u \in (0,1]} \left| \int_{-2b}^{2b} (\widehat{\varphi}_1(ux) - \widehat{\varphi}_1(0)) \cdot \widehat{\varphi}_2(x) e^{-4\pi i am x} \,  dx \right| \ll \frac{1}{m^2},
\]
where the implicit constants are independent of $m$. 
\end{lemma}

\begin{lemma}
\label{Lemma_VdC2}
For all $u > 0$ and $m \neq 0$,
\[
\left| \int_{-2b}^{2b} L_j(x;u,am) \widehat{\varphi}_2(x) e^{-4\pi i am x} \,  \dd x \right| \ll \frac{1}{u m^2},  \quad j = 1,2,
\]
where the implicit constants are independent of $m$ and $u$.
\end{lemma}

\begin{proof}[Proof of Lemma \ref{Lemma_Lowerbndeta2} assuming Lemmas \ref{Lemma_Hmu}, \ref{Lemma_Ftheta}, \ref{Lemma_VdC1} and \ref{Lemma_VdC2}]

\hfill \break
\vspace{0.2cm}

\noindent By Lemma \ref{Lemma_Hmu} it suffices to show that
\begin{equation}
\label{NeedForLowerbndeta2}
u \cdot \int_{-2b}^{2b} F_{am}(x;u) e^{-4\pi i am x} \,  dx 
= \left(\sum_{n=-\infty}^\infty (am+n) \cdot \varphi_1\left(\frac{am + n}{u}\right)\right) \cdot \varphi_2'(2am) +O\left(\frac{u}{m^2}\right).
\end{equation}
By Lemma \ref{Lemma_Ftheta} (applied with $\theta = am$) we have 
\begin{align*}
u \cdot \int_{-2b}^{2b} F_{am}(x;u) e^{-4\pi i am x} \,  dx
&=
\left( 2\pi i \int_{-2b}^{2b} x \cdot \widehat{\varphi}_2(x) e^{-4\pi i am x} \,  dx \right) \cdot \\[0.2cm]
&\cdot \sum_{n=-\infty}^\infty (am+n) \cdot \varphi_1\left(\frac{am + n}{u}\right) \\[0.3cm]
&+
u \cdot \int_{-2b}^{2b} (\widehat{\varphi}_1(ux) - \widehat{\varphi}_1(0)) \,  \widehat{\varphi}_2(x) \,  e^{-4\pi i am x} \,  dx \\[0.3cm]
&- \sum_{j=1}^2 u(-iu)^{1+j}
\int_{-2b}^{2b} L_j(x;u,am) \widehat{\varphi}_2(x) e^{-4\pi i am x} \,  dx.
\end{align*}
Note that
\[
2\pi i \int_{-2b}^{2b} x \cdot \widehat{\varphi}_2(x) e^{-4\pi i am x} \,  dx
= -\varphi_2'(2am),  \quad \textrm{for all $m$}.
\]
By Lemma \ref{Lemma_VdC1}, 
\[
\left|u \cdot \int_{-2b}^{2b} (\widehat{\varphi}_1(ux) - \widehat{\varphi}_1(0)) \,  \widehat{\varphi}_2(x) \,  e^{-4\pi i am x} \,  dx\right| = O\left(\frac{u}{m^2}\right),
\]
and by Lemma \ref{Lemma_VdC2},
\[
\left| \sum_{j=1}^2 u(-iu)^{1+j}
\int_{-2b}^{2b} L_j(x;u,am) \widehat{\varphi}_2(x) e^{-4\pi i am x} \,  dx \right| = O\left(\frac{u^2}{m^2}\right).
\]
Since $u \in (0,1)$,  the last $O$-term can be absorbed by the first $O$-term,  and
thus we have proved \eqref{NeedForLowerbndeta2}.
\end{proof}
We are thus left with the proof of the four lemmas.

\subsection{Proof of Lemmas \ref{Lemma_Hmu} and \ref{Lemma_Ftheta}}
Lemmas \ref{Lemma_Hmu} and \ref{Lemma_Ftheta} are both applications of the Poisson summation formula:

\begin{proof}[Proof of Lemma \ref{Lemma_Hmu}]
We recall from \eqref{Def_Hmu} that
\[
H_m(u) = u \cdot 
\sum_{k=-\infty}^{\infty} e^{-2\pi i am k} \cdot \int_{-2b}^{2b} \widehat{\varphi}_1(u(x+k)) \,  \widehat{\varphi}_2(x) e^{-4\pi i am x} \,  \dd x,  
\]
for all $u > 0$ and $m \neq 0$.  Hence, 
\begin{align*}
H_m(u) 
&= u \cdot \left( \sum_{k=-\infty}^\infty e^{-2\pi i am k} \,  \widehat{\varphi}_1(uk) \right) \cdot \int_{-2b}^{2b} \widehat{\varphi}_2(x) e^{-4\pi i am x} \,  \dd x \\[0.2cm]
&+ u \cdot 
\int_{-2b}^{2b} \Big( \underbrace{\sum_{k=-\infty}^\infty e^{-2\pi i amk} \left(\widehat{\varphi}_1(u(x+k)) - \widehat{\varphi}_1(uk) \right)}_{= F_{am}(x;u)}\Big) \cdot \widehat{\varphi}_2(x) e^{-4\pi i am x} \,  \dd x
\end{align*}
By Poisson's summation formula,  and since $\varphi$ is even,  
\[
u \cdot \sum_{k=-\infty}^\infty e^{-2\pi i am k} \,  \widehat{\varphi}_1(uk)
 = \sum_{n=-\infty}^\infty \varphi_1\left(\frac{am + n}{u}\right).
\]
Furthermore,  since $\supp(\widehat{\varphi}_2) \subset [-2b,2b]$,  we have
\[
\int_{-2b}^{2b} \widehat{\varphi}_2(x) e^{-4\pi i am x} \,  \dd x
= 
\int_{-\infty}^{\infty} \widehat{\varphi}_2(x) e^{-4\pi i am x} \,  \dd x = \varphi_2(2am),
\]
for all $m$,  which finishes the proof.
\end{proof}
The argument for Lemma \ref{Lemma_Ftheta} is more involved, and we need the following lemma to bound the functions $G_1$ and $G_2$:

\begin{lemma}
\label{Lemma_rho}
Let $\rho : \bR \ra [0,\infty)$ be a bounded function such 
\[
\rho(u) \ll  |u|^{-2},  \quad \textrm{for all $|u| \geq 1$},
\]
Let $M > 0$.  Then,  for all $u \in (0,1)$,  
\[
\sup_{|w| \leq M} \,  \sum_{n=1}^\infty \rho(u(n + w)) \ll_{M,\rho} \frac{1}{u},  
\]
where the implicit constants only depend on $M$ and $\|\rho\|_\infty$.
\end{lemma}

\begin{proof}
Fix $u \in (0,1)$ and $|w| \leq M$ and define
\[
S_{-} := \{ n \in \bN \,  : \,  u |n + w| \leq 1 \}
\]
and 
\[
S_j := \{ n \in \bN \,  : \,  2^{j} < u |n+w| \leq 2^{j+1} \},  \quad \textrm{for $j \geq 0$}.
\]
Then,
\begin{equation}\label{sum}
 \sum_{n=1}^\infty \rho(u(n + w)) \leq \rho(0) + \sum_{n \in S_{-}} \rho(u(n + w))
 +  \sum_{j=0}^\infty \left( \sum_{n \in S_j} \rho(u(n + w)) \right).
\end{equation}
Since $|w| \leq M$,  we see that
\[
|S_{-}| \leq \frac{1}{u} + M \qand |S_j| \leq \frac{2^{j+1}}{u} + M,  \enskip 
\textrm{for $j \geq 0$}.
\]
Hence,
\[
\sum_{n \in S_{-}} \rho(u(n + w)) \leq |S_{-}|  \cdot \|\rho\|_\infty \leq  \left(\frac{1}{u} + M\right) \cdot \|\rho\|_\infty,
\]
and
\[
\sum_{n \in S_j} \rho(u(n + w)) \ll  |S_j| \cdot 4^{-j} \leq 
\left( \frac{2^{j+1}}{u} + M\right) \cdot 4^{-j}
\]
Upon summing over all $j$,  and plugging the resulting estimates into \eqref{sum},  we are done.
\end{proof}

\begin{proof}[Proof of Lemma \ref{Lemma_Ftheta}]
Since $\varphi_1$ is even and real-valued,  so is $\widehat{\varphi}_1$.  If we take out the term corresponding to
$k=0$ in the sum defining $F_\theta(x;u)$ and split the remaining sum into real and imaginary parts, 
we get
\begin{align*}
F_\theta(x;u) 
&=
\widehat{\varphi}_1(ux) - \widehat{\varphi}_1(0) \\[0.2cm]
&+\sum_{k=1}^\infty
 \left(\widehat{\varphi}_1(u(k+x)) - 2 \cdot \widehat{\varphi}_1(uk) + \widehat{\varphi}_1(u(k-x))\right) \cdot \cos(2\pi \theta k) \\[0.2cm]
 &-i \cdot \sum_{k=1}^\infty  
 \left(\widehat{\varphi}_1(u(k+x)) - \widehat{\varphi}_1(u(k-x)) \right) \cdot \sin(2\pi \theta k).
\end{align*}
Note that $\widehat{\varphi}_1$ is smooth,  and thus
\[
\widehat{\varphi}_1(u(k+x)) - 2 \cdot \widehat{\varphi}_1(u k) + \widehat{\varphi}_1(u(k-x)) 
= 
u^2 \cdot \int_{0}^x \int_{-y}^y \widehat{\varphi}_1''(u(k + z)) \,  dz \,  dy
\]
and
\begin{align*}
\widehat{\varphi}_1(u (k+x)) - \widehat{\varphi}_1(u(k-x)) 
&= 
u \cdot \int_{-x}^x \widehat{\varphi}_1'(u (k+y)) \,  dy \\[0.2cm]
&=u \cdot \int_0^x \left(  \widehat{\varphi}_1'(u (k+y)) +  \widehat{\varphi}_1'(u (k-y)) \right) \,  dy \\[0.2cm]
&=u \cdot 
 \int_0^x \left(  \widehat{\varphi}_1'(u (k+y)) - 2 \cdot \widehat{\varphi}_1'(uk)+  \widehat{\varphi}_1'(u (k-y)) \right) \,  dy \\[0.3cm]
 &+2 u x \cdot \widehat{\varphi}'_1(uk) \\[0.2cm]
 &=
 u^3 \cdot \int_0^x \int_0^y \int_{-z}^z \widehat{\varphi}'''_1(u(k+w)) \,  dw \,  dz \,  dy+ 
 2 u x \cdot \widehat{\varphi}'_1(uk),
\end{align*}
for all $x \geq 0$,  and similarly for $x < 0$.  We conclude that
\vspace{0.1cm}
\begin{align*}
F_\theta(x;u)
&= \widehat{\varphi}_1(ux) - \widehat{\varphi}_1(0) \\[0.2cm]
&+u^2 \cdot \int_0^x \int_{-y}^y \left( \sum_{k=1}^\infty \widehat{\varphi}_1''(u(k+z)) \cdot \cos(2\pi \theta k) \right) \,  dz \,  dy \\[0.2cm]
&-i u^3 \cdot \int_0^x \int_{0}^y \int_{-z}^z 
\left( \sum_{k=1}^\infty \widehat{\varphi}_1'''(u(k + w))\cdot \sin(2\pi \theta k)   \right) \,  dw \,  dz \,  dy \\[0.2cm]
&-x \cdot iu \cdot \sum_{k=-\infty}^\infty \widehat{\varphi}_1'(uk) \,  \sin(2\pi \theta k).
\end{align*}
By Poisson's summation formula,  and since $\varphi$ is even, 
\begin{align*}
\sum_{k=-\infty}^\infty \widehat{\varphi}_1'(uk) \,  \sin(2\pi \theta k)
&= \frac{\pi}{u^2} \cdot \sum_{n=-\infty}^\infty \left( (n-\theta)\cdot \varphi_1\left(\frac{n-\theta}{u} \right) - (n+\theta)\cdot \varphi_1\left(\frac{n+\theta}{u} \right) \right) \\[0.2cm]
&= -\frac{2\pi}{u^2} \sum_{n=-\infty}^\infty (n+\theta)\cdot \varphi_1\left(\frac{n+\theta}{u} \right).
\end{align*}
Define
\[
G_1(z;u,\theta) = \sum_{k=1}^\infty \widehat{\varphi}_1''(u(k+z)) \cdot \cos(2\pi \theta k)
\]
and
\[
G_2(z;u,\theta) = \sum_{k=1}^\infty \widehat{\varphi}_1'''(u(k + w))\cdot \sin(2\pi \theta k).
\]
Since $|\widehat{\varphi}_1^{(p)}(x)| \ll_p x^{-2}$ for all $p \geq 0$,  these sums converge uniformly for fixed $u$ and $\theta$,  and thus define continuous (hence bounded) functions on $[-2b,2b]$. To see that these functions actually satisfy the bounds specified in the lemma we apply Lemma \ref{Lemma_rho} with
$M := 2b$ and $\rho := \max(|\widehat{\varphi}_1'',|\widehat{\varphi}_1'''|)$. If we now define $L_1$ and $L_2$ in terms of $G_1$ and $G_2$ as stated in the lemma, then we can write
\begin{align*}
F_{\theta}(x;u) &= x \cdot \frac{2\pi i}{u} \cdot \sum_{n=-\infty}^\infty (n+\theta) \cdot \varphi_1\left(\frac{\theta + n}{u}\right) + \widehat{\varphi}_1(ux) - \widehat{\varphi}_1(0) \\[0.2cm]
&+ u^2 \cdot L_1(x;u,\theta)  - i u^3 \cdot L_2(x;u,\theta),
\end{align*}
for all $x \in [-2b,2b]$,  which finishes the proof.
\end{proof}

\subsection{Proofs of Lemma \ref{Lemma_VdC1} and Lemma \ref{Lemma_VdC2}}

The two remaining lemmas are a consequence of the following standard result can be obtained by applying partial integration twice.

\begin{lemma}
\label{Lemma_VdC}
Let $g$ be piecewise twice continuously differentiable function
with compact support,  and suppose that $F \in C^2(\supp(g))$.  
Then,  for all real $\lambda$,  
\[
\left| \int_{-\infty}^\infty F(t) \cdot g(t) \cdot e^{i \lambda t}\,  dt \right|
\leq \frac{3}{|\lambda|^2} \cdot \max\{ \|F^{(j)}\|_\infty \|g^{(2-j)}\|_\infty \,  : \,  j = 0,1,2\},
\]
where the $\|\cdot\|_\infty$-norms are restricted to $C(\supp(g) \setminus Q)$,  where
$Q$ denotes the end points of the intervals on which $g$ is twice continuously differentiable.  
\end{lemma}

\begin{proof}[Proof of Lemma \ref{Lemma_VdC1}]
For a fixed $u \in (0,1)$,  apply Lemma \ref{Lemma_VdC} to 
\[
F = \widehat{\varphi}_1(ux) - \widehat{\varphi}_1(0) \qand 
g = \widehat{\varphi}_2 \qand \lambda = -4\pi am.
\]
Then $F$ is smooth,  $\supp(g) = [-2b,2b]$ and $\|F^{(j)}\|_\infty \ll 1$ for $j=0,1,2$,  with implicit constants independent of $u$.  
\end{proof}

\begin{proof}[Proof of Lemma \ref{Lemma_VdC2}]
Since 
\[
L_1(x;u,\theta) = \int_0^x \int_{-y}^y G_1(z;u,\theta) \,  \dd z \,  \dd y
\qand
L_2(x;u,\theta) = \int_0^x \int_0^y \int_{-z}^z G_2(w;u,\theta) \,  \dd w \,  \dd z \,  \dd y,
\]
and both $G_1(\cdot;u,\theta)$ and $G_2(\cdot;u,\theta)$ are continuous (hence bounded) on $[-2b,2b]$,  we see that both $L_1(\cdot;u,\theta)$ and $L_2(\cdot;u,\theta)$ are twice continuously differentiable on $[-2b,2b]$,  and 
\[
\|L_k^{(j)}(\cdot;u,\theta)\|_\infty \ll_b \|G_k(\cdot;u,\theta)\|_\infty,  \quad \textrm{for $j=0,1,2$ and $k=1,2$},
\]
where the $\|\cdot\|_\infty$-norms are restricted to the interval $[-2b,2b]$.  By
Lemma \ref{Lemma_Ftheta},  
\[
\|G_k(\cdot;u,\theta)\|_\infty \ll u^{-1},  \quad \textrm{for $k=1,2$},
\]
so by Lemma \ref{Lemma_VdC},  applied with $F = L_k(\cdot;u,\theta)$ and $g = \widehat{\varphi}_2$ and $\lambda = 4\pi am$,  we have
\[
\left| \int_{-2b}^{2b} L_k(x;u,\theta) \widehat{\varphi}_2(x) e^{-4\pi i am x} \,  dx
\right| \ll (m^2u)^{-1},
\]
for all $m \neq 0$,  which finishes the proof.
\end{proof}

\section{Quasicrystals which are hyperuniform}\label{Criterion}

\subsection{Hyperuniformity from repellence}
We now study cut-and-project-processes $p(\R^{d_1}, \R^{d_2}, \Gamma, W)$ as in Example \ref{ExCuP} for arbitrary dimension parameters $d_1$ and $d_2$. For such processes we provide a sufficient condition for hyperuniformity along the lines of Theorem \ref{SufficientConditionIntro}, based on the notion of repellence of lattices (cf.\ Definition \ref{DefRep}).
As demanded in Example \ref{ExCuP} we will always assume that $\Gamma$ projects injectively to $\R^{d_1}$ and densely to $\R^{d_2}$. As for $W$, we are going to assume that $W$ is Fourier smooth in the sense of Remark \ref{DecayWindows}. We then have the following criterion, which will be established in Subsection \ref{SubsecSufficientProof} below.
\begin{theorem}[Sufficient condition for hyperuniformity]\label{SufficientCondition} Let $p = p(\R^{d_1}, \R^{d_2}, \Gamma, W)$, where $W \subset \bR^{d_2}$ is Fourier smooth with exponent $\vartheta$. Assume that the dual lattice $\Gamma^{\perp}$ of $\Gamma$ is $\beta$-repellent on the right for some $\beta > 0$. Then for all sufficiently small $\varepsilon > 0$ we have
\begin{equation}\label{QuantHyp}
\widehat{\eta}_p({B}_\eps) \ll \eps^{\beta{(d_2+\vartheta)}}, \quad \text{and hence} \quad \frac{\widehat{\eta}_p({B}_\eps)}{{\Vol}_{d_1}(B_\eps)} \ll \eps^{{\beta(d_2+\vartheta)} - d_1}
\end{equation}
In particular, $p$ is hyperuniform provided that ${\beta} > \frac{d_1}{{d_2 + \vartheta}}$.
\end{theorem}
Since Euclidean balls are Fourier smooth with exponent $1$, this contains Theorem \ref{SufficientConditionIntro} from the introduction as a special case.
Before we turn to the proof of the criterion, we apply it to establish more general versions of Theorem \ref{ArithmeticIntro} and Theorem \ref{GenericIntro} from the introduction.

\subsection{Hyperuniformity of arithmetic cut-and-project processes}
We first consider the case of cut-and-project processes  $p = p(\R^{d_1}, \R^{d_2}, \Gamma, W)$, for which the dual $\Gamma^\perp$ of the underlying lattice is arithmetic in the following sense.
\begin{example}\label{ExArithmetic} Let $K$ be a totally real number field of degree $d$ with ring of integers $\cO_K$ and field embeddings $\sigma_1, \dots, \sigma_d: K \to \R$. Let us assume that $\Gamma^\perp$ is given by the image of $\cO_K$ under the diagonal embedding $\sigma_1 \times \dots \times \sigma_d: \cO_K \to \R^d$; we then say that $\Gamma^\perp$ is \emph{arithmetic}. If $(\xi_1, \dots, \xi_d)\in \Gamma^\perp \setminus\{0\}$, then $\xi_1 \cdots \xi_d \in \Z$ and hence $|\xi_1 \cdots \xi_d| \geq 1$. If we write $d = d_1 + d_2$ as the sum of two natural numbers and assume that $|\xi_1|, ..., |\xi_{d_1}| < \varepsilon$, then by the arithmetic-geometric mean inequality we have
\begin{eqnarray*}
\|(\xi_{d_1 + 1}, \dots, \xi_{d_1 + d_2})\|_\infty &\gg& \|(\xi_{d_1 + 1}, \dots, \xi_{d_1 + d_2})\|_1 \quad \geq\quad d_2 \cdot |\xi_{d_1 + 1} \cdots \xi_{d_1+d_2}|^\frac{1}{d_2}\\  &=& d_2 \cdot |\xi_1|^{-\frac 1{d_2}} \cdots |\xi_{d_1}|^{-\frac 1{d_2}} |\xi_1 \cdots \xi_{d}|^{\frac 1 {d_2}} \quad \geq \quad d_2 \cdot \varepsilon^{-\frac{d_1}{d_2}}.
\end{eqnarray*}
Thus $\Gamma^\perp$ is $\beta$-repellent on the right for $\beta := \frac{d_1}{d_2}$. 
\end{example}
\begin{corollary}[Hyperuniformity from arithmeticity]\label{HUArithmetic} Let $p = p(\R^{d_1}, \R^{d_2}, \Gamma, W)$, where $W \subset \bR^{d_2}$ is Fourier smooth with exponent $\vartheta > 0$. Assume that $\Gamma^\perp$ is an arithmetic lattice as above. Then $p$ is hyperuniform with
\[
\frac{\widehat{\eta}_p(B_\eps)}{\Vol_{d_1}(B_\eps)} \ll \eps^{\frac{d_1 \vartheta}{d_2}}.
\]
\end{corollary}
\begin{proof} Since $\Gamma^\perp$ is $\beta$-repellent with $\beta := \frac{d_1}{d_2}$, this follows from Theorem \ref{SufficientCondition} and the inequality
\[
\beta({d_2 + \vartheta}) - d_1=  \frac{d_1(d_2+\vartheta)-d_1d_2}{d_2} = \frac{d_1 \vartheta}{d_2} > 0. \qedhere
\]
\end{proof}
Since Euclidean balls are Fourier smooth with exponent $1$, this specializes to Theorem \ref{ArithmeticIntro} from the introduction. The main interest in Corollary \ref{HUArithmetic} lies in the fact that is provides \emph{explicit} examples of hyperuniform cut-and-project processes. However, the underlying lattices are of a very special kind. To complement Corollary \ref{HUArithmetic} we are thus going to consider \emph{generic} lattices (with respect to the unique invariant measure class on the space of all lattices) in the next subsection.

\subsection{Hyperuniformity of generic model sets}
In this subsection we establish the following genericity result for cut-and-project processes with Fourier smooth windows.
\begin{theorem}[Hyperuniformity for generic cut-and-project processes]\label{GenericHyperuniformity} Let $W \subset \bR^{d_2}$ be Fourier smooth. Then for almost every lattice $\Gamma$ the cut-and-project process $p = p(\R^{d_1}, \R^{d_2}, \Gamma, W)$ is hyperuniform. More precisely, if $W$ is Fourier smooth with exponent $\vartheta$, then for every $\delta > 0$ and  almost every $\Gamma$ we have
\[
\frac{\widehat{\eta}_p({B}_\eps)}{{\Vol}_{d_1}(B_\eps)} \ll_\delta \eps^\frac{d_1(\vartheta-\delta)}{d_2+\delta}.
\]
\end{theorem}
For Euclidean balls, Theorem \ref{GenericHyperuniformity} specializes to Theorem \ref{GenericIntro} from the introduction. We now turn to the proof. Throughout, let $\Gamma$ and $W$ be as in Theorem \ref{SufficientCondition}. We write
\begin{equation}\label{GammaperpMatrix}
\Gamma^\perp= 
\left(
\begin{matrix}
A & B \\
C & D
\end{matrix}
\right) \bZ^{d_1 + d_2}
\end{equation}
for matrices $A,B,C,D$ of appropriate sizes. We can then express $\beta$-repellence in terms of these matrices.
\begin{definition}
A real $(d_1 \times d_2)$-matrix $E$ is \emph{$\alpha$-repellent} if 
there exists $Q_E>0$ such that
\[
\|p + Eq\|_\infty \geq \|q\|_\infty^{-\alpha}  \quad \textrm{for all $(p,q) \in \bZ^{d_1}  \times \bZ^{d_2}$ such that $\|q\|_\infty \geq Q_E$}.
\]
\end{definition}
\begin{lemma}\label{LemmaRepel}
Let $\Gamma^\perp$ be given by \eqref{GammaperpMatrix} and assume that
\begin{equation}\label{DetConditions}
\det(A) \neq 0 \qand \det(D-CA^{-1}B) \neq 0.
\end{equation}
If $A^{-1}B$ is $\alpha$-repellent,  then $\Gamma^\perp$ is $\beta$-repellent on the right,  
for every $\beta < \frac 1 \alpha$.
\end{lemma}

\begin{proof} By assumption, every $\xi = (\xi_1, \xi_2) \in \Gamma^\perp$ can be written as
\[
(\xi_1,\xi_2) = (Ap + Bq,Cp+Dq)  \quad \textrm{for some $(p,q) \in \bZ^{d_1} \times \bZ^{d_2}$}.
\]
Let $\eps > 0$,  and suppose that
\[
\|\xi_1\|_\infty = \|A(p + A^{-1}Bq)\|_\infty < \eps.
\]
Since $\det(A) \neq 0$ we have $\|p + A^{-1}Bq\|_\infty \ll \eps$, and since $A^{-1}B$ is $\alpha$-repellent,  we further have
\[
\|q\|_\infty^{-\alpha} \leq \|p + A^{-1}Bq\|_\infty \ll \eps,  \quad \textrm{for all $p \in \bZ^{d_1}$ and $\|q\|_\infty \gg 1$}.
\]
In particular, for all sufficiently small $\eps$ we have
\[
\|q\|_\infty \gg \eps^{-1/\alpha}.
\]
Finally,  note that
\begin{align*}
\|\xi_2\|_\infty 
&=
\|Cp + Dq\|_\infty  = \|CA^{-1}\lambda_1 + (D-CA^{-1}B)q\|_\infty \\[0.2cm]
&\gg
\|(D-CA^{-1}B)q\|_\infty -  \eps.
\end{align*}
Since $\det((D-CA^{-1}B) \neq 0$ we have
\[
\|(D-CA^{-1}B)q\|_\infty \gg \|q\|_\infty \gg \eps^{-1/\alpha}.
\]
We conclude
that if $\eps$ is sufficiently small (so that $\eps^{-1/\alpha} \gg \eps$), then
\[
\|\xi_2\|_\infty \geq \eps^{-\beta} \quad \text{for all }\beta < \frac 1 \alpha.
\]
In particular, $\Gamma^\perp$ is $\beta$-repellent on the right. 
\end{proof}
\begin{corollary} If $W \subset \bR^{d_2}$ is Fourier smooth with exponent $\vartheta$ and $\Gamma^\perp$ is given by \eqref{GammaperpMatrix} with $A,B,C,D$ satisyfing \eqref{DetConditions} and $A^{-1}B$ is $\alpha$-repellent for some $\alpha < \frac{d_2}{d_1} + \delta$ with $\delta \in (0, \vartheta)$, then $p := p(\R^{d_1}, \R^{d_2}, \Gamma, W)$ satisfies
\[
\frac{\widehat{\eta}_p({B}_\eps)}{{\Vol}_{d_1}(B_\eps)} \ll_\delta \eps^\frac{d_1(\vartheta-\delta)}{d_2+\delta}.
\]
In particular, if $W$ is Fourier smooth and $A^{-1}B$ is $\alpha$-repellent for some $\alpha < \frac{d_2}{d_1} + \vartheta$, then  $p(\R^{d_1}, \R^{d_2}, \Gamma, W)$ is hyperuniform.
\end{corollary}
\begin{proof} If we set $\beta := \frac{d_1}{d_2 + \delta}$, then it follows from Lemma \ref{LemmaRepel} that $\Gamma^\perp$ is $\beta$-repellent on the right. We then dedue from \eqref{QuantHyp} that
\[
\frac{\widehat{\eta}_p({B}_\eps)}{{\Vol}_{d_1}(B_\eps)} \ll \eps^{\beta({d_2+\vartheta}) - d_1} = \eps^{d_1\left(\frac{d_2+\vartheta}{d_2+\delta} -1\right)} = \eps^\frac{d_1(\vartheta-\delta)}{d_2+\delta}.
\]
Since
\[
\beta({d_2 + \vartheta}) = \frac{d_1(d_2+\vartheta)}{d_2+\delta} > d_1,
\]
the corollary follows from Theorem \ref{SufficientCondition}.
\end{proof}
We have thus reduced the proof of Theorem \ref{GenericHyperuniformity} to showing that the set 
\begin{equation*}
\left\{
g=
\left(
\begin{matrix}
A & B \\
C & D
\end{matrix} 
\right) \in \GL_{d_1 + d_2}(\bR) \,  : \,  \det(A),  \det(D-CA^{-1}B) \neq 0,  \enskip \textrm{$A^{-1}B$ is $\alpha$-repellent} \right\},
\end{equation*}
or equivalently (since $\det(g) = \det(A) \det(D-CA^{-1}B)$ if $\det(A) \neq 0$), 
that the set
\begin{equation}\label{Calpha}
\cC_\alpha = \left\{
\left(
\begin{matrix}
A & B \\
C & D
\end{matrix} 
\right) \in \GL_{d_1 + d_2}(\bR) \,  : \,  \det(A) \neq 0,  \enskip \textrm{$A^{-1}B$ is $\alpha$-repellent} \right\}
\end{equation}
is conull with respect to the Haar measure class for all $\alpha > \frac{d_2}{d_1}$. For this we use following consequence of the Khintchine-Groshev theorem:
\begin{lemma}[Khintchine-Groshev]\label{KhinchinGroshev} For every $\alpha > \frac{d_2}{d_1}$ the set
\[
\widehat{\cC}_\alpha := \{E \in \R^{d_1 \times d_2} \mid E \text{ is $\alpha$-repellent}\}
\]
is conull with respect to Lebesgue measure class.
\end{lemma}
\begin{proof} We apply the version of the Khintchine-Groshev theorem on p.\ 2 of \cite{BV} with $\psi(x) := x^{-\alpha}$. Since $d_2-1-\alpha d_1 < -1$ we have
\[
\sum_{q=1}^\infty q^{d_2-1}\psi(q)^{d_1} = \sum_{q=1}^\infty q^{d_2-1-\alpha d_1} < \infty,
\]
hence the theorem states (in our notation) that
\[
\cE_\alpha := \{E \in \R^{d_1 \times d_2}  \mid \|(p,E q)\|_\infty < \|q\|^{-\alpha} \text{ holds for infinitely many }p \in \Z^{d_1}, q\in \Z^{d_2} \setminus\{0\} \}
\]
is a Lebesgue nullset in $\R^{d_1 \times d_2} $. We claim that the complement of $\cE_\alpha$ is contained in $\widehat{\cC}_\alpha$. Indeed, if $E \not \in \cE_\alpha$ and 
\[
Q_E := \max\{\|q\|_\infty \mid \|(p,E q)\|_\infty < \|q\|^{-\alpha}, p \in \Z^{d_1}, q\in \Z^{d_2} \setminus\{0\}\},
\]
then for $q \in \Z^{d_2}$ with $\|q\|_\infty > Q_E$ we have $\|p+Eq\|_\infty\geq  \|q\|^{-\alpha}$, hence $E$ is $\alpha$-repellent.
\end{proof}
The proof of the theorem now follows:
\begin{proof}[Proof of Theorem \ref{GenericHyperuniformity}] Denote by $m$ a probability measure on $\mathrm{GL}_{d}(\R)$ in the measure class of Haar measure. Note that the open set 
\[
 \GL_{d_1 + d_2}(\bR)^\sharp := \left\{\left(
\begin{matrix}
A & B \\
C & D
\end{matrix} 
\right) \in \GL_{d_1 + d_2}(\bR) \,  : \,  \det(A) \neq 0\right\} \subset  \GL_{d_1 + d_2}(\bR).
\]
is $m$-conull.  We further observe that the map
\[
\sigma:  \GL_{d_1 + d_2}(\bR)^\sharp  \to  \mathrm{Mat}_{d_1, d_2}(\R), \quad \left(
\begin{matrix}
A & B \\
C & D
\end{matrix} 
\right) \mapsto A^{-1}B\] satisfies \[\sigma\left(\left(\begin{matrix}
A & B \\
C & D
\end{matrix} \right)\left(\begin{matrix}
S& 0 \\
0 & T
\end{matrix} \right)\right) = S^{-1} \sigma\left(\left(\begin{matrix}
A & B \\
C & D
\end{matrix} \right)\right)T,
\]
and hence $\sigma_*m$ is quasi-invariant under $ \GL_{d_1}(\bR) \times \GL_{d_2}(\bR)$,  whence of Lebesgue measure class. By Lemma \ref{KhinchinGroshev} we deduce that $\widehat{\cC}_\alpha $ is conull with respect to $\sigma_*m$, hence $\sigma^{-1}(\widehat{\cC}_\alpha) = \cC_\alpha$ is conull with respect to Haar measure class.
\end{proof}


\subsection{Proof of the hyperuniformity criterion}\label{SubsecSufficientProof} 

We now turn to the proof of Theorem \ref{SufficientCondition}. We need an estimate for sums over uniformly discrete point sets.
\begin{lemma}\label{LemmaSumOverSep}
For every $d \geq 1$ and $\vartheta > 0$ there exists a constant  $C_{d,\vartheta}$ with the following property: If $R > 1$ and $\Delta_R \subset \bR^d \setminus\{0\}$ such that $\Delta_R \cup \{0\}$ is $R$-uniformly discrete, then
\[
\sum_{\delta \in \Delta_R} \max(1,\|\delta\|_\infty)^{-(d+\vartheta)} \ll_{d,\vartheta} R^{-(d+\vartheta)}.
\]
\end{lemma}
\begin{proof} Fix $R > 1$ and let $\Delta_R \cup \{0\} \subset \bR^d$ be $R$-uniformly discrete. This implies that 
\begin{equation}
\min_{\delta \in \Delta_R} \|\delta\|_\infty \geq R.
\end{equation}
We now consider the disjoint decomposition
\[
\bR^d = \bigsqcup_{k \in \bZ^d} \frac{R}{2} \cdot \left( [-1/2,1/2)^d + k\right).
\]
Intersecting with $\Delta_R$ and using that $R>1$ we obtain
\[
\Delta_R = \bigsqcup_{\|k\|_\infty \geq 2}  \underbrace{\left(\Delta_R \cap \frac{R}{2} \cdot \left( [-1/2,1/2)^d + k\right) \right)}_{=: \Delta_R(k)},
\]
where for each $k$ with $\|k\|_\infty \geq 2$ there is at most one point in the intersection $\Delta_R(k)$.  Also note that if $\delta_k \in \Delta_R(k)$,  then by the triangle inequality
\[
\min\{1,\|\delta_k\|_\infty\} \geq R(\|k\|_\infty-1/2).
\]
Hence,
\begin{align*}
\sum_{\delta \in \Delta_R} \max\{1,\|\delta\|_\infty^{-(d+\vartheta)}\} 
&\leq 
\sum_{\|k\|_\infty \geq 2} \left(R(\|k\|_\infty-1/2)\right)^{-(d+\vartheta)} \\[0.2cm]
&\ll_d 
R^{-(d+\vartheta)} \sum_{l =2}^\infty \frac{l^{d-1}}{(l-1/2)^{d+\vartheta}} \\[0.2cm]
&\ll_d
R^{-(d+\vartheta)} \cdot \sum_{l=2}^\infty l^{-(1+\vartheta)}, 
\end{align*}
where the implicit constants only depend on $d$.  Since $\vartheta > 0$,  the sum on the right hand side converges to a constant depending only on $\vartheta$.
\end{proof}
To apply this lemma we observe:
\begin{lemma}\label{LemmaDeltaSep} Let $\Gamma$ as in Theorem \ref{SufficientCondition} and let $\eps \in (0,1)$. If
\[
\Delta := \left\{ \xi_2 \,  : \,  (\xi_1,\xi_2) \in \Gamma^{\perp} \setminus \{(0,0)\},
\enskip \|\xi_1\|_\infty < \eps \right\},
\]
then $\Delta \cup \{0\}$ is $(2\eps)^{-\beta}$-uniformly discrete.
\end{lemma}
\begin{proof} Since $\Gamma^{\perp}$ is $\beta$-repellent on the right,  we have 
\[
\|\xi_2\|_\infty \geq \eps^{-\beta} > (2\eps)^{-\beta},  \quad 
\textrm{for all $\xi_2 \in \Delta$}.
\]
Furthermore,  if
\[
(\xi_1,\xi_2) \neq  (\xi_1',\xi_2') \in \Gamma^{\perp} \setminus \{(0,0)\} \qand \|\xi_1\|_\infty,  \|\xi'\|_\infty < \eps,
\]
then
\[
(\xi_1-\xi_1',\xi_2-\xi'_2) \in \Gamma^{\perp} \setminus \{(0,0)\},  \qand \|\xi_1-\xi_1'\|_\infty < 2\eps,
\]
and thus $\|\xi_2-\xi_2'\|_\infty \geq (2\eps)^{-\beta}$. 
\end{proof}
\begin{proof}[Proof of Theorem \ref{SufficientCondition}] Assume that $W$ is Fourier smooth with exponent $\vartheta > 0$. With $\Delta$ defined as in Lemma \ref{LemmaDeltaSep} we have 
\[
\widehat{\eta}(\widehat{B}_\eps(0)) \quad
=  \quad \sum_{\xi_2 \in \Delta} |\widehat{\chi}_W(\xi_2)|^2 \quad
\ll \sum_{\xi_2 \in \Delta} \max(1,\|\xi_2\|)^{d_2 + \vartheta},
\]
an since $\Delta \sqcup \{0\}$ is $(2\eps)^{-\beta}$-uniformly discrete we may apply Lemma \ref{LemmaSumOverSep}  (with $R := (2\eps)^{-\beta}$) to obtain
\[
\widehat{\eta}(B_\eps(0)) \ll \eps^{{\beta(d_2 + \vartheta)}},
\]
for all sufficiently small $\eps > 0$.
\end{proof}


\section{Meyerian point processes with positive asymptotic number variance}\label{SecMeyerian}

\subsection{A criterion for positive asymptotic number variance of suspension processes}

\begin{construction}
Assume that $T$ is an invertible ergodic pmp transformation of a standard probability space $(Z, \cB, \theta)$ and that $B \subset Z$ is a Borel subset such that $\bigcup_n T^{-n}B = Z$. We then obtain a invariant simple point process in $\Z$ by
\[
p^\Z: (Z, \theta) \to M(\Z), \quad z \mapsto \delta_{\Lambda^\Z_z}, \quad \text{where} \quad \Lambda^\Z_z = \{n \in \Z \mid T^nz \in B\}.
\]
This is a special case of the general construction of a transverse process as discussed in Appendix \ref{AppTrans}. The point process in $p^\Z$ gives rise to an invariant hard-core simple point process in $\R$ via suspension: If we set  $\Omega := (\R \times Z)/\Z$, where $\Z$ acts on $\R \times Z$ on the right by $(t,z).n := (t-n, T^nz)$, and denote by $q: \R \times Z \to \Omega$ the canonical projection, then $\bP := q_*(\Vol_1|_{[0,1)} \otimes \theta)$ is an invariant ergodic probability measure on $\Omega$ and the suspended process is given by
\[
p: (\Omega, \bP) \to M(\R), \quad p_{[t,z]} = \delta_{\Lambda^\Z_z - t}.
\]
We refer to $p$ as the \emph{suspension process} with parameters $(Z, \theta, T, B)$. This process is again a transverse process in the sense of Appendix \ref{AppTrans}, cf.\ Example \ref{sus}, and since $\Lambda^\Z_z - t \subset \Z -t $ it is contained in a random translate of $\Z$. In particular, a suspension process is Meyerian provided $\Lambda^\Z$ is almost surely relatively dense in $\Z$. 
\end{construction}
\begin{definition} A function class $f \in L^2_o(Z,\theta)$ is called an \emph{$L^2$-coboundary} with respect to $T$ if there exists $F$ in $L^2(\theta)$ such that $f = F - F \circ T$.\end{definition}
\begin{proposition}[Coboundary criterion]\label{ConzeLeBorgneConvenient} Let $p$ be a suspension process with parameters $(Z, \theta, T, B)$. Assume that
\begin{equation}
\sum_{n=1}^\infty n \cdot | (\theta(B \cap T^{-n}B) - \theta(B)^2)| < \infty,
\end{equation}
Then the limit $\ANV(p)$ exists, and we have $\ANV(p) = 0$ if and only if the function $f_B \in L^2_o(Z)$ given by $f_B := \chi_B - \theta(B) \cdot 1$ is an $L^2$-coboundary.
\end{proposition}
In the situation of the proposition one can actually give an explicit formula for $\ANV(p)$ in terms of the function $f_B$. For this we recall that the \emph{asymptotic variance} of a function  $f \in L_o^2(Z,\theta)$ is defined as 
\[
\sigma^2(f) := \sum_{n=-\infty}^\infty \langle f \circ T^n,  f\rangle_{L^2(\theta)}.
\]
With this notation we are going to show that $\ANV(p) = \sigma^2(f_B)$. The proposition is a straight-forward consequence of a lemma of Conze and Le Borgne (cf.\ \cite[Lemma 2.2]{ClB}). Since our notation is quite different, we include the proof below.
\begin{lemma}[Conze-Le Borgne]
\label{Lemma_CLB}
Let $f \in L_o^2(Z,\theta)$ with asymptotic variance $\sigma^2(f)$ such that 
\[
C_f := \sum_{n=1}^\infty n \cdot |\langle f \circ T^n,  f\rangle_{L^2(\theta)}| < \infty.
\]
Then, 
\[
\sup_{R > 0} \left| \sum_{n=-\infty}^\infty \rho_R(n) \langle f \circ T^n, f \rangle_{L^2(\theta)} - 2R \cdot \sigma^2(f) \right|
\leq 12 C_f + \|f\|_{L^2(\theta)}^2,
\]
and if $\sigma^2(f) = 0$,  then $f$ is an $L^2$-coboundary.
\end{lemma}
\begin{proof}[Proof of Proposition \ref{ConzeLeBorgneConvenient} modulo Lemma \ref{Lemma_CLB}] Let $\Omega := (\R \times Z)/\Z$, where the right-action of $\Z$ on $\R \times Z$ is given by  $(t,z).n := (t-n, T^nz)$, and denote by $q: \R \times Z \to \Omega$, $(t,z) \mapsto [t,z] $ the canonical quotient map. We use the fact that, by Example \ref{sus}, the process $p$ is a transverse process with underlying cross section $\cT := q(\{0\} \times B)$ and corresponding Palm measure $\nu := q_*(\delta_0 \otimes \theta|_B)$. By Corollary \ref{VarPalm} we thus have
\begin{eqnarray*}\Var_p(B_R) &=&  \left(\sum_{\lambda \in \Lambda_\cT} \rho_R(\lambda) \cdot \nu(\cT \cap \lambda^{-1}.\cT)\right) - \nu(\cT)^2 \cdot \Vol_1(\rho_R)\\
&=&  \left(\sum_{n=-\infty}^\infty \rho_R(n) \cdot  \theta(B \cap T^{-n}B)\right) - \theta(B)^2 \cdot (2R)^2\\
&=&  \underset{:=S_1(R)}{\underbrace{\left(\sum_{n=-\infty}^\infty \rho_R(n) (\theta(B \cap T^{-n}B) - \theta(B)^2)\right)}}-\theta(B)^2\cdot\underset{:=S_2(R)}{\underbrace{\left(\left( \sum_{n=-\infty}^\infty \rho_R(n)\right)-(2R)^2\right)}}
\end{eqnarray*}
where $\rho_R(t) = (\chi_{B_R} \ast \chi_{B_R})(t) =  (2R - |t|) \chi_{[-2R,2R]}(t)$, and hence
\begin{equation}\label{ANVS1S2}
\ANV(p) = \lim_{R \to \infty} \frac{\Var_p(B_R)}{\Vol_1(B_R)} = \lim_{R \to \infty} \frac{S_1(R)}{2R}  + \theta(B)^2 \cdot \frac{S_2(R)}{2R}
\end{equation}
By the Poisson summation formula we have
\[
S_2(R) = R^2 \cdot \sum_{m \neq 0} \left( \frac{\sin(2\pi R m)}{\pi R m} \right)^2
=  \sum_{m \neq 0} \frac{\sin^2(2\pi m R)}{\pi^2 m^2},
\]
which is bounded in $R$, hence the second summand in \eqref{ANVS1S2} vanishes.  On the other hand we can write the first summand as
\[
\frac{S_1(R)}{2R} = \frac{1}{2R}\sum_{n=-\infty}^\infty \rho_R(n) \langle f_B \circ T^n, f_B \rangle_{L^2(\theta)},
\]
hence we deduce from the first part of Lemma \ref{Lemma_CLB} that $\ANV(p) = \sigma^2(f_B)$. In particular, $\ANV(p)$ exists, and by the second part of the lemma its vanishing implies that $f_B$ is an $L^2$-coboundary. Conversely, $\sigma^2$ vanishes on all $L^2$-coboundaries, hence the proposition holds.
\end{proof}
\begin{proof}[Proof of Lemma \ref{Lemma_CLB}] 
Let $N$ be a positive integer and define 
\[
S_N f = \sum_{n=0}^N f \circ T^n.
\]
Note that
\begin{align*}
\|S_N f\|_{L^2(\theta)}^2 
&= \sum_{m,n=0}^N \langle f \circ T^{m-n},  f \rangle_{L^2(\theta)}
= \sum_{|n| \leq N} \left(N+1 - |n|\right) \cdot \langle f \circ T^n,  f \rangle_{L^2(\theta)},
\end{align*}
hence,
\begin{align*}
\|S_N f\|^2 - (N+1) \cdot \sigma^2(f)
&= - \sum_{|n| \leq N} |n| \cdot \langle f \circ T^n ,f \rangle_{L^2(\theta)} \\[0.2cm]
&-(N+1) \cdot \sum_{|n| > N} \langle f \circ T^n, f \rangle_{L^2(\theta)},
\end{align*}
and thus
\[
\left| \|S_N f\|^2 - (N+1) \cdot \sigma^2(f) \right|
\leq 2 \cdot \sum_{n=1}^\infty n \cdot |\langle f \circ T^n,f \rangle_{L^2(\theta)}| = 2C_f.
\]
Fix $R > 0$,  and write $R = N_R + u_R$ for unique $N_R \in \bN_o$ and $u_R \in [0,1)$.  Then,  from the formula for $\|S_N f\|^2_{L^2(\theta)}$ above,  
\begin{align*}
\sum_{n=-\infty}^\infty \rho_R(n) \cdot \langle f \circ T^n ,f \rangle_{L^2(\theta)} 
&= \sum_{|n| \leq 2R} (2R-|n|)  \cdot \langle f \circ T^n ,f \rangle_{L^2(\theta)} \\[0.2cm]
&=
\sum_{|n| \leq 2N_R} (2N_R-|n|)  \cdot \langle f \circ T^n ,f \rangle_{L^2(\theta)} \\[0.2cm]
&+2(R-N_R) \cdot \sum_{|n| \leq 2N_R}  \langle f \circ T^n ,f \rangle_{L^2(\theta)} \\[0.2cm]
&+\sum_{2N_R < |n| \leq 2R} (2R-|n|) \cdot \langle f \circ T^n ,f \rangle_{L^2(\theta)} 
\\[0.2cm]
&= \|S_{2 N_R} f\|^2_{L^2(\theta)} + L(R),
\end{align*}
where
\[
L(R) = 2(R-N_R) \cdot \sum_{|n| \leq 2N_R}  \langle f \circ T^n ,f \rangle_{L^2(\theta)} 
+\sum_{2N_R < |n| \leq 2R} (2R-|n|) \cdot \langle f \circ T^n ,f \rangle_{L^2(\theta)}.
\]
Note that
\[
|L(R)| \leq 2 \cdot C_f + 6 \cdot C_f = 8 \cdot C_f,
\]
since there are at most 3 terms in the last sum.  We conclude that
\begin{align*}
\sum_{n=-\infty}^\infty \rho_R(n) \cdot \langle f \circ T^n ,f \rangle_{L^2(\theta)}
- 2R \cdot \sigma(f)
&= 
\sum_{n=-\infty}^\infty \rho_R(n) \cdot \langle f \circ T^n ,f \rangle_{L^2(\theta)}
- \|S_{2N_R} f\|^2_{L^2(\theta)} \\[0.2cm]
&+
 \|S_{2N_R} f\|^2_{L^2(\theta)} - (2N_R +1) \cdot \sigma^2(f) \\[0.4cm]
 &+
 (2N_R + 1 - 2R) \cdot \sigma^2(f).
\end{align*}
Since $\sigma^2(f) \leq \|f\|_{L^2(\theta)}^2 + 2 C_f$,  we see from above that
\[
|\sum_{n=-\infty}^\infty \rho_R(n) \cdot \langle f \circ T^n ,f \rangle_{L^2(\theta)}
- 2R \cdot \sigma(f)|
\leq 12 C_f + \|f\|_{L^2(\theta)}^2,
\]
uniformly in $R$.  Finally,  note that if $\sigma^2(f) = 0$,  then from the inequalities above,  we see that $\|S_N f\|^2_{L^2(\theta)}$
is a bounded sequence.  Hence,  if we set 
\[
F_M := \frac{1}{M} \sum_{N=1}^M S_M f,
\]
then there is a sub-sequence $(M_k)$ such that $(F_{M_k})$ weakly converges to an element $F$ in $L^2(\theta)$.  Furthermore,  since
\[
S_N f \circ T = S_N f - f + f \circ T^{N+1}
\]
\[
F - F \circ T = \lim_k (F_{M_k}  - F_{M_k} \circ T ) = \lim_k \left(f - \frac{1}{M_k} \sum_{N=1}^{M_k} f \circ T^{N+1}\right).
\]
Since $T$ is ergodic and $f \in L^2_o(Z,\theta)$,  the last averages tend to zero in $L^2(Z,\theta)$ by the mean ergodic theorem,  and thus $F - F \circ T = f$,  showing
that $f$ is an $L^2$-coboundary.
\end{proof}
To summarize: 
\begin{remark}\label{6ToShow}
Assume that  $p$ is a suspension process with parameters $(Z, \theta, T, B)$ such that
\[
\sum_{n=1}^\infty n \cdot | (\theta(B \cap T^{-n}B) - \theta(B)^2)| < \infty,
\]
and $ f_B := \chi_B - \theta(B) \cdot 1 \in L^2_o(Z, \theta)$ is \emph{not} an $L^2$-coboundary with respect to $T$. Then $p$ has positive asymptotic number variance and in particular is not hyperuniform.
\end{remark}
\subsection{The main construction}
The main idea in order to construct a suspension process with positive asymptotic number variance is to choose the parameters $(Z, \theta, T, B)$ as a mixing system. A typical example of a mixing transformation is given by the ``$\times 2$''-map on the circle $\bT = \Z \backslash \R$, i.e. by the map $T_o([x]) := [2x]$. However, since the continuous endomorphisms $T_o : \bT \to \bT$ is not invertible, we will have to work with an invertible ($2$-adic) extension. 

Denote by $\Q_2$ (respectively $\Z_2$) the field of $2$-adic numbers (respectively ring of $2$-adic integers). Every $y \in \bQ_2$ can be written uniquely as $y = m + k/2^N$ for some $m \in \bZ_2$ and for some $N \geq 1$ and odd integer $k$, and we write $\{y\}_2 := k/2^N$. We now consider the compact abelian groups
\[
K:= (\R \times \Q_2)/\Z[1/2] \qand M := (\{0\} \times \Z_2) + \Z[1/2] < K
\]
We equip these groups with the corresponding Haar probability measures $m_K$ and $m_M$ respectively.
We observe that the map $M\backslash K \cong \bT$, $[(x,y)] + M \mapsto x-\{y\}_2 + \Z$ is an isomorphism of topological groups, and denote by
\[
\pi: K \to M\backslash K \cong \bT, [x,y] \mapsto x-\{y\}_2 + \Z
\]
the canonical projection. Then the endomorphism $T_o$ of $\bT$ lifts via $\pi$ to a continuous automorphism $T \in \Aut({K})$ given by $T([x,y]) := [2x, 2y]$ such that the diagram
\[\begin{xy}
\xymatrix{
K \ar[d]_\pi  \ar[r]^T & K\ar[d]^\pi\\
\bT \ar[r]^{T_o} & \bT
}
\end{xy}\]
commutes. Given $q \in (1/2, 3/4]$ we now define subsets 
\[
B_o^{(q)} := ([0, 1/2] \cup(q,1)) + \Z \subset \bT \qand B^{(q)} := \pi^{-1}(B_o^{(q)}) \subset K.
\]
We observe that
\begin{equation}\label{AqBig}
B_o^{(q)} \cup T_o^{-1}(B_o^{(q)}) = \bT, \quad \text{and hence} \quad B^{(q)} \cup T^{-1}(B^{(q)}) = K.
\end{equation}
For any $q \in (1/2, 3/4]$ we consider the suspension process $p^{(q)}$ with parameters $(K, m_K, T, B^{(q)})$. 
It follows from \eqref{AqBig} that for all $q \in (1/2, 3/4)$ and all $x \in K$ the set
\[
\Lambda^\Z_x = \{n \in \Z \mid T^nx \in B^{(q)}\}
\]
satisfies $\Lambda^\Z_x \cup (\Lambda^\Z_x + 1) = \bZ$ and thus $\Lambda^\Z_x$ is $2$-syndetic in $\Z$. We are going to show the following result, which in particular implies Theorem \ref{PosANVIntro} from the introduction.
\begin{theorem}[Quasicrystals with positive asymptotic number variance]\label{PosANV} There exists a dense subset $S \subset (1/2, 3/4]$ such that for all $q \in S$ the quadruple $(K, m_K, T, B^{(q)})$ satisfies the conditions of Remark \ref{6ToShow}. In particular, for all $q \in S$ the asymptotic number variance $\mathrm{ANV}\left(p^{(q)}\right)$ exists and is strictly positive.
\end{theorem}
\subsection{Proof of Theorem \ref{PosANV}}
Given $q \in (1/2, 3/4]$ we define
\[
\psi^{(q)}_{o} = \chi_{B_o^{(q)}} - m_{\bT}(B_o^{(q)}) \qand \psi^{(q)} = \psi^{(q)}_o \circ \pi = \chi_{B^{(q)}} - m_{\bT}(B^{(q)}).
\]
According to Remark \ref{6ToShow} the following two lemmas combine to prove Theorem \ref{PosANV}.
\begin{lemma}\label{Lemma1} For every $q \in (1/2, 3/4]$ we have
\begin{equation}\label{EqLemma1}
\sum_{n=1}^\infty n \cdot | (m_K(B^{(q)} \cap T^{-n}B^{(q)}) - m_K(B^{(q)})^2)| = \sum_{n=1}^\infty n |\langle \psi^{(q)} \circ T^n,\psi^{(q)} \rangle| < \infty.
\end{equation}
\end{lemma}
\begin{lemma}\label{Lemma2}
There is a dense subset $S \subset (1/2,3/4]$ such that for every $q \in S$ the
function $\psi^{(q)} \in L^2(K)$ is not an $L^2$-coboundary for $T$.
\end{lemma}
We now turn to the proof of these lemmas. 
\begin{proof}[Proof of Lemma \ref{Lemma1}] The first equality in \eqref{EqLemma1} holds by definition. For the second equality, 
fix $q \in (1/2,3/4]$ and set $\psi_o := \psi_o^{(q)}$ and $\psi := \psi^{(q)}$. We observe that $\psi_o$ is a function of bounded variation on $\bT$ with zero integral, hence by \cite[Proposition 3.3.14]{Gr} we have
\begin{equation}
\label{fourierdecay}
|\widehat{\psi}_o(k)| \ll \frac{1}{|k|}  \quad \textrm{for all $k \neq 0$}.
\end{equation}
Now for all $n \geq 0$ we have
\[
\int_K \psi(T^{n}(x)) \overline{\psi(x)} \,  \dd m_K(x)
\;=\;
\int_{\bT} \psi_o(T_o^{n}(x)) \overline{\psi_o(x)} \,  \dd m_{\bT}(x) \\[0.2cm]
\;=\;
\sum_{k \neq 0} \widehat{\psi}_o(2^n k) \overline{\psi_o(k)}.
\]
Applying Cauchy-Schwartz and using \eqref{fourierdecay} twice we thus find for all $n \geq 0$,
\begin{eqnarray*}
 |\langle \psi^{(q)} \circ T^n,\psi^{(q)} \rangle| &=& \left|\int_K \psi(T^{n}(x)) \overline{\psi(x)} \,  \dd m_K(x) \right|\\ &\leq& \left( \sum_{k \neq 0} |\widehat{\psi}_o(2^n k)|^2 \right)^{1/2} 
\left( \sum_{k \neq 0} |\widehat{\psi}_o(k)|^2 \right)^{1/2}\\
&\leq& \left(\sum_{k \neq 0} \frac{|2^n k \cdot \widehat{\psi}_o(2^n k)|^2}{2^{2n} k^2} \right)^{1/2} \left( \sum_{k \neq 0} \frac 1 {k^2} \right)^{1/2}\\
&\ll&\left(2^{-2n}\cdot\sum_{k \neq 0} \frac{1}{ k^2} \right)^{1/2} \quad \ll \quad 2^{-n}.
\end{eqnarray*}
The lemma then follows by multiplying by $n$ and summing over $n$.
\end{proof}
For the proof of Lemma \ref{Lemma2} we need to find a condition that ensures that some $M$-invariant function $\psi$ (in our case the function $\psi^{(q)}$ for certain values of $q$) is \emph{not} a coboundary. This is provided by the following lemma. Here we denote by $T^{*} \in \Aut(\widehat{K})$ the automorphism of the dual group $\widehat{K}$ given by
\[
T^{*}(\xi)(k) = \xi(T^{-1}(k)),  \quad \textrm{for $\xi \in \widehat{K}$ 
and $k \in K$}.
\]
We also denote by $M^\perp \subset \widehat{K}$ the annihilator of $M$ and note that $T(M) \subsetneq M$.
\begin{lemma}\label{xidagger}
Assume that $\psi$ is an $M$-invariant $L^2$-coboundary with respect to $T$.  Suppose that there exists $\xi^\dagger \in M^\perp$ such that
\begin{enumerate}
\item $\displaystyle \lim_{n \ra \pm \infty} T^{*n}(\xi^\dagger) = \infty$ (i.e.\ the sequence $(T^{*n}(\xi^\dagger))$ leaves any finite subset of $\widehat{K}$) \vspace{0.2cm}
\item $T^{*n}(\xi^\dagger) \notin M^\perp$,  for all $n \geq 1$.
\end{enumerate}
\vspace{0.2cm}
Then,
\[
\sum_{n=-\infty}^0 |\widehat{\psi}(T^{*n}(\xi^\dagger))| < \infty
\implies 
\sum_{n=-\infty}^0 \widehat{\psi}(T^{*n}(\xi^\dagger)) = 0.
\]
\end{lemma}
\begin{proof} By assumption there exists $\varphi \in L^2(K)$
such that
\[
\psi = \varphi - \varphi \circ T.
\]
We deduce that for all $\xi \in \widehat{K}$ and $n \in \bZ$ we have
\begin{equation}\label{DualConv}
\widehat{\psi}(T^{*n}(\xi)) = \widehat{\varphi}(T^{*n}(\xi)) -  \widehat{\varphi}(T^{*{(n+1)}}(\xi)).
\end{equation}
Now fix $\xi^\dagger \in \widehat{K}$ satisfying the Conditions (1) and (2) of the lemma and define
\[
\beta_n := \widehat{\psi}(T^{*n}(\xi^\dagger)) \qand \gamma_n :=  \widehat{\varphi}(T^{*n}(\xi^\dagger)).
\]
By \eqref{DualConv} we then have $\beta_n = \gamma_n - \gamma_{n+1}$ for all $n \in \Z$. Since $\varphi \in L^2(K)$ we have $\widehat{\varphi} \in c_o(\widehat{K})$. It thus follows from (1) that 
\begin{equation}\label{gamma0}
\lim_{n \ra \pm \infty} \gamma_n = 0.
\end{equation}
Now since $\psi$ is $M$-invariant, we have $\widehat{\psi}(\xi) = 0$ for all $\xi \notin M^{\perp}$. By (2) we thus have $\widehat{\psi}(T^{*n}(\xi^\dagger)) = 0$ for all $n \geq 1$. Thus if $\sum_{n=-\infty}^0 |\beta_n|$ converges, then $(\beta_n)$ is absolutely summable. Since $\beta_n=\gamma_n - \gamma_{n+1}$ we then deduce from \eqref{gamma0} that \[\sum_{n=-\infty}^0 \beta_n = 0.\qedhere\]
\end{proof}
\begin{remark}\label{xidaggerexplicit}
We are going to apply Lemma \ref{xidagger} to the function
\[
\xi^\dagger(x,y) = e^{2\pi i(x - \{y\}_2)},  \quad \textrm{for $[x,y] \in K$}.
\]
This is well-defined, since $e^{2\pi i(x - \{y\}_2)}$ vanishes on the diagonally embedded $\Z[1/2]$, and satisfies $\xi^\dagger|_M = 1$, i.e.\  $\xi^\dagger \in M^{\perp}$. We claim that $\xi^\dagger$ satisfies Properties (1) and (2) of Lemma \ref{xidagger}. Concerning (1) we first observe that if we define
\[
\alpha(x) := e^{2\pi i \{y\}_2},  \enskip y \in \bQ_2 \qand \beta(x) := e^{-2\pi i y},  \enskip x \in \bR,
\]
and if we set $\alpha_x(\cdot) = \alpha(x \cdot)$ and $\beta_y(\cdot) = \beta(y \cdot)$, then we obtain an isomorphism
\[
\R \times \Q_2 \to \widehat{\R \times \Q_2}, \quad (x,y) \mapsto \beta_x \otimes \alpha_y,
\]
and $\widehat{K}$ can be identified with a discrete subset of $\R \times \Q_2$. Given $[x,y] \in K$ we have
\[
(T^{*n}\xi_o)([x,y]) =  e^{2\pi i(2^{-n}x - \{2^{-n}y\}_2)},
\]
which under the above identification corresponds to the sequence $(2^{-n}x, 2^{-n}y) \in \R\times \Q_2$. Now if $n \to -\infty$ then the first component leaves any compact subset of $\R$, and if $n \to \infty$, then the second component leaves any compact subset of $\Q_2$, hence $\xi_o$ satisfies (1). Furthermore, for all $n \geq 1$ and every odd integer $m$ we have 
\[
(T^{*n}\xi_o)(0,m) = \xi_o(0,2^{-n}m) = e^{-2\pi i m/2^n} \neq 1,
\]
thus $T^{*n}(\xi_o) \notin M^{\perp}$, which is (2).
\end{remark}

\begin{proof}[Proof of Lemma \ref{Lemma2}] We define $\xi^\dagger \in M^\perp$ as in Remark \ref{xidaggerexplicit}. Given $q \in (1/2, 3/4]$ we then set $\beta_n^{(q)} := \widehat{\psi}^{(q)}(T^{*n}(\xi^\dagger))$. By Lemma \ref{xidagger} and Remark \ref{xidaggerexplicit} it then suffices to show that for all $q$ in a dense subset $S \subset (1/2, 3/4)$ we have
\begin{equation}\label{Lemma21}
\sum_{n=-\infty}^0 |\beta_n^{(q)}| < \infty \qand
\sum_{n=-\infty}^0 \beta_n^{(q)} \neq 0.
\end{equation}
To describe the coefficients $\beta_{-n}^{(q)}$ for $n \geq 0$ we denote by $\xi^\dagger_o \in \widehat{M \backslash K}$ the unique character such that $\xi^\dagger = \xi^\dagger_o \circ \pi$. Then for all $n \geq 0$ we have
\begin{align*}
\beta_{-n}^{(q)} &= \int_K \psi^{(q)}(x) (T^{*(-n)}\xi^\dagger)(x) \,  dm_K(x)
= \int_K \psi^{(q)}(x) \xi^\dagger(T^{n}(x)) \,  dm_K(x) \\[0.2cm]
&= \int_{M \backslash K} \psi^{(q)}_o(Mx) \xi^\dagger_o(T_o^{n}(Mx)) \,  
dm_{K \backslash M}(Mx).
\end{align*}
In view of the explicit form of $\xi^\dagger$ and $T_o$ this implies that $\beta_{-n}^{(q)} = \widehat{\psi}^{(q)}_o(2^{-n})$, and hence we have to show that
for all $q$ in a dense subset $S \subset (1/2, 3/4)$ we have
\begin{equation}\label{Lemma22}
\sum_{n=0}^\infty | \widehat{\psi}^{(q)}_o(2^{-n})| < \infty \qand
\sum_{n=0}^\infty  \widehat{\psi}^{(q)}_o(2^{-n}) \neq 0.
\end{equation}
The first condition actually holds, by the same argument as above, for all $q \in (1/2, 3/4)$,  since $\psi^{(q)}_o$ is of bounded variation. Now assume for contradiction that the second condition of \eqref{Lemma22} does not hold on a dense subset $S \subset (1/2, 3/4)$. We then find an open interval $I$ in the complement of $S$ such that
\begin{equation}\label{pretheta}
\sum_{n=0}^\infty  \widehat{\psi}^{(q)}_o(2^{-n}) = 0 \quad \text{for all }q\in I.
\end{equation}
We now define a periodic function 
\[
\theta: \R \to \C, \quad \theta(q) :=  \sum_{n=0}^\infty \frac{e^{-2\pi i 2^n q}}{2^n}.
\]
We then deduce from \eqref{pretheta} and the fact that
\[
 \widehat{\psi}^{(q)}_o(2^{-n}) =  \widehat{\chi}_{[0,1/2]}(2^n) + \widehat{\chi}_{(q,1)}(2^n) =  \widehat{\chi}_{[0,1/2]}(2^n) + \frac{1}{2\pi i} \left( \frac{e^{-2\pi i 2^n q}}{2^n} - \frac{1}{2^n} \right),
\]
that $\theta(q)$ is constant on $I$, hence in particular differentiable on $I$. Since $\theta$ is given by 
a lacunary Fourier series, it then follows from a classical result of Hardy \cite[Proposition 3.6.2]{Gr} that
\[
2^n \cdot \widehat{\theta}(-2^n) \ra 0. 
\]
This contradicts the fact that, by the explicit formula, $\widehat{\theta}(-2^n) = \frac{1}{2^n}$.
\end{proof}

\section{Number rigidity}\label{SecNR}

\subsection{An abstract rigidity criterion}
Let $p$ be a locally square-integrable random measure on $\bR^d$. Given a Borel subset $B \subset \bR^d$ we denote by $B^c$ its complement.
\begin{definition}\label{rigstat} A Borel function $f: \R^d \to \R$ is called a \emph{rigidity statistics} for $p$ if for every bounded Borel set $B \subset \bR^d$ there exists a conull set  $\Omega_B \subset \Omega$ such that for all $\omega \in \Omega_B$ the integral
\[
\int_{B} f \,  \dd p_\omega
\]
only depends on the restriction $p_\omega|_{B^c}$.
\end{definition}
In particular, a point process $p$ is \emph{number rigid} as defined in the introduction if and only if the constant function $f = 1$ is a rigidity statistics in the sense of Definition \ref{rigstat}. We will need the following rigidity criterion, which has some history. An early version of it was established by 
\cite[Theorem 6.1]{GP}.  We refer to \cite{C} for an informative survey of rigidity for point processes.
\begin{lemma}
Let $p :(\Omega, \bP) \to M(\R^d)$ be a locally square-integrable random measure on $\bR^d$ and let $f : \bR^d \ra \bR$ be a Borel function.  Suppose that there exists a sequence $(f_n)$ of bounded Borel functions such that
\begin{enumerate}[(i)]
\item $\cP f_n \in L^2(\Omega, \bP)$ for every $n$.  
\item $f_n \ra f$ pointwise as $n \ra \infty$. 
\item $\Var_p({f}_n) \ra 0$ as $n \ra \infty$.
\end{enumerate}
Then $f$ is a rigidity statistics for $p$.
\end{lemma}
\begin{proof} Let $c_n := \bE[\cP f_n]$. Since $\Var_p(f_n) = \Var(\cP f_n) = \bE[(\cP f_n - c_n)^2] = \|\cP f_n - c_n\|^2_{L^2(\Omega, \bP)}$, we deduce from (iii) that $\cP f_n - c_n$ converges to $0$ in norm in $L^2(\Omega, \bP)$. We thus find a subsequence $(n_k)$ and a conull subset $\Omega_o \subset \Omega$ such that
\[
p_\omega(f_{n_k}) - c_{n_k} \to 0\quad \text{for all}\; \omega \in \Omega_o.
\]
Now for every bounded Borel set $B \subset \bR^d$ and every $\omega \in \Omega_o$ we obtain
\begin{align*}
\int_B f \,  \dd p_\omega
&= \lim_{k \ra \infty} \int_B f_{n_k} \,  \dd p_\omega = \lim_{k \ra \infty} 
\left( p_\omega(f_{n_k}) - c_{n_k}\right) +   \lim_{k \ra \infty} \left(c_{n_k} - \int_{B^c} f_{n_k} \,  \dd p_\omega \right)\\
&= \lim_{k \ra \infty} 
\left(c_{n_k} - \int_{B^c} f_{n_k} \,  \dd p_\omega \right).
\end{align*}
We may thus set $\Omega_B := \Omega_o$ for all bounded Borel sets $B$ and since the right-hand side only depends on $p_\omega|_{B^c}$, $f$ is a rigidity statistics for $p$.
\end{proof}
Specializing to the case $f := 1$ we obtain:
\begin{corollary}\label{NRCrit1} A locally square-integrable random measure  $p :(\Omega, \bP) \to M(\R^d)$ on $\bR^d$ is number rigid if there exists a sequence $(f_n)$ of bounded Borel functions such that
\begin{enumerate}[(i)]
\item $\cP f_n \in L^2(\Omega, \bP)$ for every $n$.  
\item $f_n \ra 1$ pointwise as $n \ra \infty$. 
\item $\Var_p({f}_n) \ra 0$ as $n \ra \infty$.\qed
\end{enumerate}
\end{corollary}
Based on this criterion, we can now establish our spectral criterion for number rigidity, which is Lemma \ref{NRCritIntro} from the introduction.
\begin{lemma}\label{NRCrit}
If there exists a sequence $\eps_n \searrow 0$ such that
\begin{equation}\label{epsn}
\widehat{\eta}_p(B_{\eps_n}) \ll \eps_n^{2d+\delta},
\end{equation}
for some $\delta > 0$, then $p$ is number rigid.
\end{lemma}
\begin{proof}
We want to construct a sequence $(f_n)$ of Schwartz functions on $\bR^d$ satisfying Conditions (i) -- (iii) of Corollary \ref{NRCrit1}.  For this let $\phi(u) = e^{-\pi u^2}$ and $\phi_t(x) = \phi(t \|x\|)$,  for $t > 0$
and $x \in \bR^d$.  Then,  $\phi_t$ is a Schwarz function on $\bR^d$ and $\phi_t(x) \ra 1$ for all $x$ as $t \ra 0$, hence if $t_n \searrow 0$, then the sequence $f_n := \phi_{t_n}$ satisfies Conditions (i) and (ii) of Corollary \ref{NRCrit1}. We are going to show that (iii) holds for some specific choice of sequence $t_n \searrow 0$. To find such a sequence $t_n$ we observe that
\[
\widehat{\phi}_t(\xi) = t^{-d} \phi(\|\xi\|/t),  \quad \textrm{for $\xi \in \bR^d$},
\]
and thus
\begin{eqnarray*}
\Var_p(\phi_t)
&=& \int_{\bR^d} |\widehat{\phi}_t(\xi)|^2 \,  \dd \widehat{\eta}_p(\xi) \quad = \quad  \int_{\bR^d}  t^{-2d} \phi(\|\xi\|/t)^2 \,  \dd \widehat{\eta}_p(\xi)\\
&=& t^{-2d} \int_{0}^\infty \widehat{\eta}_p(\{\xi \mid \phi(\|\xi\|/t) \geq \sqrt u\})\, \dd u.
\end{eqnarray*}
Since $\phi$ is monotone decreasing on $[0,\infty)$ we have
\[
\phi(\|\xi\|/t) \geq \sqrt u \iff \|\xi\| \leq w := t \phi^{-1}(\sqrt u) \iff \xi \in B_w
\]
Since $u = \phi(w/t)^2$ we have $\dd u = 2\phi(w/t)\phi'(w/t)t^{-1} \dd w$ and thus a substitution yields
\[
\Var_p(\phi_t) = t^{-(2d+1)} \cdot \int_0^\infty  \widehat{\eta}_p(B_w) 2 \phi(w/t)|\phi'(w/t)| \, \dd w.
\]
Now let $\eps_n$ as in \eqref{epsn} and choose $\gamma$ with $0<\gamma < \frac{\delta}{2d+1}$. If we set $t_n := \eps_n^{1+\gamma}$, then $t_n \searrow 0$ and the functions
$f_n := \varphi_{t_n}$ satisfy
\begin{align*}
\Var_p(f_n) 
&\leq t_n^{-(2d+1)} \cdot \int_{0}^{\eps_n}  \widehat{\eta}_p(B_w) 2 \phi(w/t_n)|\phi'(w/t_n)| \, \dd w\\[0.2cm]
&+ 
t_n^{-(2d+1)} \cdot \int_{\eps_n} ^\infty
 \widehat{\eta}_p(B_w) 2 \phi(w/t_n)|\phi'(w/t_n)| \, \dd w \\[0.2cm]
&\ll t_n^{-(2d+1)} \cdot \eps_n \cdot \widehat{\eta}_\mu(B_{\eps_n}) \cdot \|\phi\|_\infty\|\phi'\|_\infty \\[0.2cm]
&+ 
t_n^{-(2d+1)} \cdot \int_{\eps_n}^\infty w^{d} \cdot \phi(w/t_n) |\phi'(w/t_n)| \,  \dd w \\[0.2cm]
&\ll t_n^{-(2d+1)} \cdot \eps_n^{2d+1 + \delta} +  \int_{\eps_n/t_n}^\infty \left(\frac{u}{t_n}\right)^d \cdot \phi(u) |\phi'(u)| \,  \dd u\\[0.2cm]
&\ll t_n^{\delta - \gamma(2s+1)} + \int_{\eps_n^{-\gamma}}^\infty \left({u}{\eps_n^{-(\gamma+1)}}\right)^d \cdot \phi(u) |\phi'(u)| \,  du.
\end{align*}
The first term clearly tends to zero as $n \ra \infty$.  Concerning the second term,  we observe that
\[
\int_{\eps_n^{-\gamma}}^\infty  u^d \cdot \phi(u) |\phi'(u)| \,  \dd u
\leq \left( \int_{\eps_n^{-\gamma}} u^d |\phi(u)|^2 \,  \dd u \right)^{1/2}
\cdot \left( \int_{\eps_n^{-\gamma}} u^d |\phi'(u)|^2 \,  \dd u \right)^{1/2},
\]
and,  since $\varphi(u) = e^{-\pi u^2}$,  
\[
\int_{\eps_n^{-\gamma}}^\infty  u^d \cdot |\phi^{(j)}(u)| \,  du \ll e^{-\pi \eps_n^{-\gamma}},  \quad \textrm{for $j=1,2$},
\]
where the implicit constants are independent of $n$.  Since 
$\eps_n^{-d(\gamma+1)} e^{-\pi \eps_n^{-\gamma}} \ra 0$ as $n \ra \infty$,  we conclude that the second term above also tends to zero,  which finishes the proof.
\end{proof}

\section{A $p$-adic cut-and-project process}\label{SecStealth}

\begin{example}[A $p$-adic cut-and-project process]\label{Expadic} For any fixed prime number $p$ we consider the cut-and-project process $p = p(G, H, \Gamma, W)$ in $\Q_p$ with parameters
\[
G = \bQ_p,\quad H = \bR, \quad \Gamma = \{ (\gamma,\gamma) \in G \times H \,  : 
\gamma \in \bZ[1/p] \} \qand W = [-1/2,1/2].
\]
Note
that $\cF = \bZ_p \times [0,1)$ is a fundamental domain for $\Gamma$,  so if we normalize the Haar measures so that $m_{\bR}([0,1)) = 1$ and $m_{\bQ_p}(\bZ_p) = 1$, then the covolume of $\Gamma$ is equal to one. 
\end{example}
To compute the diffraction of this process we need to parametrize the Pontryagin dual of $L := G \times H$. 
\begin{remark}\label{DualL}
Every $y \in \Q_p$ can be written uniquely as $x = m + k/p^N$, where $m \in \bZ_p$, $N \geq 1$ and $k$ is an integer coprime to $p$. We then set $\{x\}_p := k/p^N$ and define characters on $\Q_p$ and $\R$ respectively by
\[
\alpha(x) := e^{2\pi i \{x\}_p},  \enskip x \in \bQ_p \qand \beta(y) := e^{-2\pi i y},  \enskip y \in \bR,
\]
and if we set $\alpha_x(\cdot) = \alpha(x \cdot)$ and $\beta_y(\cdot) = \beta(y \cdot)$, then we obtain an isomorphism
\[
L \to \widehat{L}, \quad (x,y) \mapsto \alpha_x \otimes \beta_y.
\]
If we use this isomorphism to identify $L$ with $\widehat{L}$, then our lattice $\Gamma$ becomes self-dual in the sense that
$\Gamma^\perp = \{ (x,y) \in L \,  : \,  (\alpha_x \otimes \beta_y)|_\Gamma = 1 \} = \Gamma$.
\end{remark}
In view of Remark \ref{DualL}, Theorem \ref{DiffractionFormula} specializes as follows to the case at hand:
\begin{corollary}\label{DiffPAdic} The centered diffraction of the cut-and-project process $p$ from Example \ref{Expadic} is given by
\[
\pushQED{\qed}\widehat{\eta}_p =  \sum_{\gamma \in \bZ[1/p] \setminus \{0\}} |\widehat{\chi}_W(\gamma)|^2 \,  \delta_\gamma.\qedhere\popQED
\]
\end{corollary}
From this we deduce the following result, which implies Theorem \ref{stealthIntro} from the introduction.
\begin{corollary}[Stealth]\label{stealth} The centered diffraction of the cut-and-project process $p$ from Example \ref{Expadic} vanishes on $\Z_p$. In particular, $p$ is stealthy.
\end{corollary}
\begin{proof} Under our standing identification of $\widehat{\R}$ with $\R$ we have
\[
\widehat{\chi}_W(y) = \int_{-1/2}^{1/2} e^{2\pi i y t} \,  \dd t = \frac{\sin(\pi y)}{\pi y},  \quad y \in \bR.
\]
Since $\bZ_p \cap \bZ[1/p] = \bZ$ we deduce from Corollary \ref{DiffPAdic} that
\[
\widehat{\eta}_\mu(\bZ_p) = \sum_{n \neq 0} \left(\frac{\sin(\pi n)}{\pi n} \right)^2 = 0. \qedhere
\]
\end{proof}

\newpage
\appendix

\section{Poisson vs.\ cut-and-project diffraction}\label{DiffFormulas}
In this appendix we collect the well-known proofs of the diffraction formulae from Proposition \ref{DiffPoiss} and Theorem \ref{DiffractionFormula}.
\subsection{Poisson diffraction}
The proof of Proposition \ref{DiffPoiss} is based on the following lemma.
\begin{lemma} If $(Y,m)$ is a $\sigma$-finite measure space and $p$ is an $m$-Poisson process on $Y$, then $\Var_p(f) = m_G(|f|^2)$ for all $f \in L^2(Y, m)$.
\end{lemma}
\begin{proof} It suffices to show the formula for \emph{real}-valued functions $f$. By definition, the variance measure $\Var_p$ satisfies 
\[
\Var_p(f) = \Var(\cP f) = \bE[(\cP f)^2] - \bE[\cP f]^2,
\]
where $\cP f$ denotes the linear statistics of $f$, hence it suffices to show that, for an $m$-Possion measure $\mu$ and a non-negative bounded real-valued function $f$,
\begin{equation}\label{Poiss1}
\bE[\cP f] = \int_Y f \,  dm \qand  \bE[(\cP f)^2] = \int_Y f^2 \,  dm + \left(\int_Y f \,  dm\right)^2.
\end{equation}
By a standard approximation argument it suffices to establish \eqref{Poiss1} in the case where 
\begin{equation}\label{Poissf}
f = \sum_{k=1}^r c_k \chi_{B_k} \quad (B_k \subset Y \text{ Borel}, c_k \in \R).
\end{equation}
Now, by Property (i) of a Poisson measure, we have
\[\bE[\cP \chi_B] = \sum_{k=0}^\infty k \cdot 
\mu\left(\left\{ p \,  : \,  p(B) = k \right\}\right) = \sum_{k=0}^\infty 
\frac{k \cdot m(B)^k \,  e^{-m(B)}}{k!} = m(B),
\]
for any bounded Borel set $B \subset Y$, and similarly
\[
\bE[(\cP \chi_B)^2] = \sum_{k=0}^\infty k^2 \cdot 
\mu\left(\left\{ p \,  : \,  p(B) = k \right\}\right) = \sum_{k=0}^\infty 
\frac{k^2 \cdot m(B)^k \,  e^{-m(B)}}{k!} = m(B)(m(B)+1).
\]
For $f$ as in \eqref{Poiss1} we deduce from linearity of the expectation that
\[
\bE[\cP f] = \sum_{k=1}^r c_k m(B_k) = \int_Y f \,  dm,
\]
and using Property (ii) of a Poisson measure we obtain
\begin{align*}
 \bE[(\cP f)^2] 
&= \sum_{k,l=1}^r c_k c_l \, \bE[(\cP \chi_{B_k})\cdot (\cP \chi_{B_l})] = \sum_{k=1}^r c_k^2 \cdot m(B_k)(m(B_k) + 1) \\
&+ \sum_{k \neq l} c_k c_l \cdot m(B_k) m(B_l) = \int_Y f^2 \,  dm + \left( \int_Y f \,  dm \right)^2.\qedhere
\end{align*}
\end{proof}
\begin{proof}[Proof of Proposition \ref{DiffPoiss}] Specializing the lemma to the case $(Y, m) = (G, m_G)$ we see that
\[
\eta_p(f \ast f^*) = \Var_p(f) = \int_G |f|^2 \,  dm_G = (f \ast f^*)(e),
\]
which implies (i), and then (ii) follows by taking Fourier transforms.
\end{proof}
 
 \subsection{Cut-and-project diffraction}
We now turn to the proof of Theorem \ref{DiffractionFormula}; we keep the notation of the theorem. Given a bounded Borel function $F: L \to \C$ with bounded support we may define its $\Gamma$-periodization
\[
 \Per_\Gamma(F): \Omega \to \C, \quad \Per_\Gamma(F)(\Gamma g) = \sum_{\gamma \in \Gamma} F(\gamma g),
\]
which is bounded on $\Omega$ and hence in $L^2(\Omega, \bP)$. Note that for every bounded Borel function $F$ on $L$ with bounded support we have
\begin{equation}\label{PerInt}
\int_L F \,  dm_L = \int_{\cF} \Per_\Gamma(F) \,  dm_L
= m_L(\cF) \cdot \bE[\Per_\Gamma(F)],
\end{equation}
\begin{lemma}\label{Parseval} For any bounded Borel function $F: L \to \C$ with bounded support we have
\[
\|\Per_\Gamma(F)\|^2_{L^2(\Omega, \bP)} = \frac{1}{m_L(\cF)^2} \sum_{\xi \in \Gamma^{\perp}} |\widehat{F}(\xi)|^2.
\]
\end{lemma}
\begin{proof} Every $\xi \in \Gamma^\perp$ descends to a bounded measurable function on $\Omega$, which we denote by the same letter; then $L^2(\Omega, \bP) = \widehat{\bigoplus}_{\xi \in \Gamma^{\perp}} \bC \cdot \xi$. For $\xi \in \Gamma^\perp$ we have  $\Per_\Gamma(F) \cdot \overline{\xi} = \Per_\Gamma(F \cdot \overline{\xi})$, hence \eqref{PerInt} yields
\[
\langle \Per_\Gamma(F), \xi\rangle_{L^2(\Omega, \bP)} = \frac{1}{m_L(\cF)} \int_{\cF} \Per_\Gamma(F \cdot \overline{\xi}) \,  \dd m_L = \frac{1}{m_L(\cF)} \int_L F \cdot \overline{\xi} \,  \dd m_L = \frac{\widehat{F}(\xi)}{m_L(\cF)}.
\]
Since $\Per_\Gamma(F) \in L^2(\mu)$,  we conclude from Parseval's Theorem that
\[
\|\Per_\Gamma(F)\|^2_{L^2(\Omega, \bP)} = \sum_{\xi \in \Gamma^\perp} |\langle \Per_\Gamma(F), \xi\rangle_{L^2(\Omega, \bP)}|^2 = \frac{1}{\mathrm{covol}(\Gamma)^2} \sum_{\xi \in \Gamma^{\perp}} |\widehat{F}(\xi)|^2.\qedhere
\]
\end{proof}
\begin{proof}[Proof of Theorem \ref{DiffractionFormula}] By definition, we have for every $f \in \cL_c^\infty(G)$ that
\[
\widehat{\eta}^+_p(|\widehat{f}|^2) = \eta^+_p(f \ast f^*) = M^2_p(f \otimes \bar f) = \bE[|p_\omega(f)|^2] = \int_{\Omega} \left| \delta_{\Lambda_\omega}(f)\right|^2 \dd\bP(\omega)
\]
The key observation is now that
\[
 \delta_{\Lambda_\omega}(f) = \sum_{g \in \Lambda_\omega} f(g) = \Per_\Gamma(f \otimes \chi_W)(\omega),  \quad \omega \in \Omega.
\]
With Lemma \ref{Parseval} we deduce that
\[
\widehat{\eta}^+_p(|\widehat{f}|^2) = \|\Per_\Gamma(f \otimes \chi_W)\|^2_{L^2(\Omega, \bP)}
= \frac{1}{\covol(\Gamma)^2} \sum_{(\xi_1,\xi_2) \in \Gamma^{\perp}} |\widehat{f}(\xi_1)|^2 \cdot |\widehat{\chi}_W(\xi_2)|^2.
\]
This establishes the formula for the diffraction, and the formula for the centered diffraction then follows from Proposition \ref{RemoveAtom}.
\end{proof}

\section{Transverse point processes}\label{AppTrans}

In this appendix we describe a large class of hard-core point processes called \emph{transverse point processes}, which were introduced in \cite{BHK}. It will turn out that all of the processes considered in this article, in particular cut-and-project processes, fall into this class.

\subsection{Transverse point processes and transverse measures}

We work in the following setting: We consider a pmp action of a unimodular lcsc group $G$ on a standard probability space $(\Omega, \mathcal F, \bP)$. Given a measurable subset $\cT \subset \Omega$ and an element $\omega \in \Omega$ we define subsets
\begin{equation}\label{Lambdaomega}
\Lambda_\omega := \{g \in G \mid g.\omega \in \cT\} \subset G \qand \Lambda_\cT := \bigcup_{\omega \in \cT} \Lambda_\omega \subset G;
\end{equation}
the latter is called the \emph{return time set} of $\cT$. 
\begin{definition} $\cT$ is called a \emph{cross section} if it intersects every orbit in a non-empty and at most countable set. Given an identity neighbourhood $U$ in $G$, we say that $\cT$ is \emph{$U$-separated} if $\Lambda_\cT \cap U = \{e\}$. It is called \emph{separated} if it is $U$-separated for some $U$. We say that $\cT$ is \emph{cocompact} if $\Omega = K.\cT$ for some compact subset $K \subset G$.
\end{definition}
\begin{theorem}[Conley]
For every pmp action of a unimodular lcsc group $G$ on a standard probability space $(\Omega, \mathcal F, \bP)$ there exists a cocompact separated cross section $\cT$.
\end{theorem}
For the proof see \cite[Theorem 2.4]{Slutsky}. If $\cT$ is a separated cross section (not necessarily cocompact), then we say that $(\Omega, \bP, G, \cT)$ is a \emph{transverse system}. Following \cite{BHK} we associate a transverse point process and a transverse measure with every transverse system; in probabilistic language this transverse measure is just the Palm measure of the transverse point process.
\begin{proposition}[{\cite[Lemma 3.3]{BHK}}] \label{TransverseProcess}  For every transverse system $(\Omega, \bP, G, \cT)$ the map 
\[
p: (\Omega, \bP) \to M(G), \quad \omega \mapsto p_\omega := \delta_{\Lambda_\omega},
\]
where $\Lambda_\omega$ is defined as in \eqref{Lambdaomega}, defines an invariant simple hard-core point process.
\end{proposition}
\begin{definition} The point process $p$ from Proposition \ref{TransverseProcess} is called the \emph{transverse point process} associated with the transverse system $(\Omega, \bP, G, \cT)$. 
\end{definition}
With every transverse system $(\Omega, \bP, G, \cT)$ one can also associate a transverse measure $\nu$ on $\cT$, the \emph{Palm measure} of the associated transverse process $p$ \cite{BHK, Last}.  This measure admits the following two characterizations \cite[Thm.\ 1.14]{BHK}: If $\cT$ is $U$-separated and $V \subset G$ is an identity neighbourhood with $VV^{-1} \subset U$, then 
\[
\bP(V.Y) = m_G(V) \cdot \nu(Y) \quad (Y \subset \cT \text{ Borel}).
\]
Alternatively, given a non-negative Borel function $\varphi$ on $G \times \cT$
such that $\{(g,\omega) \in G \times \cT \mid \varphi(g,\omega) > 0\}$ projects to a bounded set in $G$, we may define
\[
T\varphi(\omega) = \sum_{g \in \Lambda_\omega} \varphi(g^{-1},g.\omega),  \quad \omega \in \Omega.
\]
Then $T\varphi$ is a bounded Borel function on $\Omega$ and 
\begin{equation}\label{PalmDef}
\bE[T\varphi] = (m_G \otimes \nu)(\varphi).
\end{equation}

\subsection{Examples of transverse point processes}
The simplest kind of transverse point processes are periodic point processes:
\begin{example}[Periodic case] Let $G$ be an lcsc group and let $\Gamma$ be a lattice in $G$. Then $G$ acts on $\Omega:= \Gamma\backslash G$ via $g.\Gamma h := \Gamma hg$, and this action preserves a unique invariant probability measure $\bP$. Moreover, $(\Omega, \bP, G, \{\Gamma e\})$ is a transverse system, with Palm measure is of the form $\nu = 1/\covol(\Gamma) \cdot \delta_{\Gamma e}$ and the transverse point process is given by $p_{\Gamma g} = \delta_{\Gamma g}$.
\end{example}
It turns out that cut-and-project processes can also be seen as transverse point processes:
\begin{example}[Cut-and-project process]\label{RanMS}
Let $G$, $H$ be lcsc group, $\Gamma < G \times H$ be a lattice projecting injectively to $G$ and densely to $H$ and $W \subset H$ be a relatively compact set with non-empty interior.  We define a (minimal) action of $G$ on $\Omega := \Gamma\backslash (G \times H)$ by $g_1.(\Gamma(g,h)) := \Gamma(gg_1^{-1}, h)$; this action admits a unique invariant probability measure $\bP$ (cf. \cite{BHP1}). If we now denote by $q: G \times H \to \Omega$ the canonical projection and set $\cT := q(\{e\} \times W)$, then $(\Omega, \cP, G, \cT)$ is a transverse system with Palm measure
\[
\nu = \frac{1}{\covol(\Gamma)} \cdot q_*(\delta_e \otimes m_H|_W).
\]
and associated transverse process $p(G, H, \Gamma, W)$, the cut-and-project process with parameters $(G, H, \Gamma, W)$.
\end{example}
Our third example concerns suspensions of transverse point processes in $\Z$.
\begin{example}[Suspended transverse system] \label{sus}
We consider a transverse system $(Z, \theta, \Z, B)$, i.e. $Z$ is a standard Borel space, $\Z$ acts measurably on $Z$ by powers of an invertible measurable transformation $T: Z \to Z$ and $B \subset Z$ is a Borel set such that $\bigcup_n T^{-n}B = Z$. We then obtain a transverse system $(\Omega, \bP, \R, \cT)$ via suspension: 

We define a right-action of $\Z$ on $\R \times Z$ by $(t,z).n := (t-n, T^nz)$ and define $\Omega := (\R \times Z)/\Z$; we denote by $q: \R \times Z \to \Omega$, $(t,z) \mapsto [t,z] $ the canonical quotient map. Then $\R$ acts on $\Omega$ by $x.[t,z] := [x+t, z]$ preserving the probability measure $\bP := q_*(\Vol_1|_{[0,1)} \otimes \theta)$, and if we set $\cT := q(\{0\} \times B)$, then $(\Omega, \bP, \R, \cT)$ is a transverse system, called the suspension of $(Z, \theta, \Z, B)$.

One checks that the Palm measure of this transverse system is $\nu := q_*(\delta_0 \otimes \theta|_B)$ and that if $p^\Z: Z \to M(\Z)$ denotes the transverse process associated with $(Z, \theta, \Z, B)$ and $\Lambda^\Z := \supp(p^\Z)$, then the transverse process associated with $(\Omega, \bP, \R, \cT)$ is given by
\[
p: \Omega \to M(\R), \quad p_{[t,z]} = \delta_{\Lambda^\Z_z - t}.
\]
In particular, $\supp(p_{[t,z]}) \subset \Z - t$, i.e. $p$ is contained in a random translate of $\Z$.
\end{example}

\subsection{Auto-correlation and variance of transverse point processes}

We now derive a formula for the autocorrelation and variance of a transverse process in terms of the corresponding Palm measure. Thus let $(\Omega, \bP, G, \cT)$ be a transverse system with transverse point process $p$ and Palm measure $\nu$. Given $f \in \cL^\infty_c(G)$ we denote by
\[
\cP_\cT f: \cT \to \C, \quad \cP_\cT f(\omega) :=  \sum_{g \in \Lambda_\omega} f(g)
\]
the restriction of the linear statistic of $f$ to the cross section $\cT$.
\begin{proposition}[Palm formula for autocorrelation]\label{PalmAutocor} For any transverse system $(\Omega, \bP, G, \cT)$  with transverse process $p$ and Palm measure $\nu$ we have
\[
\eta^+_{p}(f) = \nu(\cP_\cT f) \qand \eta_p(f) = \nu(\cP_\cT f) -\nu(\cT)^2 \cdot m_G(f) \quad (f \in \cL^\infty_c(G)).
\]
\end{proposition}
\begin{proof} Fix $f \in \cL^\infty_c(G)$ and let $\rho$ be a non-negative normalized Borel function on $G$ with bounded support. We define
\[
\varphi: G \times \cT \to \C, \quad \varphi(g,\omega ) := \rho(g^{-1}) \cP_\cT f(\omega). 
\]
Since the point process $p_\omega$ is given by
\[
p_\omega(f) = \sum_{g \in \Lambda_\omega} f(g),
\]
we deduce that for $f_\rho(g_1, g_2) := f(g_1g_2^{-1})\rho(g_2)$ we have
\[
T \varphi(\omega) = \sum_{g_2 \in \Lambda_\omega} \rho(g_2) \cP_\cT f(g_2^{-1}.\omega) = \sum_{g_2 \in \Lambda_\omega} \rho(g_2) \sum_{g_1 \in \Lambda_{\omega}} f(g_1g_2^{-1})
 = (p_\omega \otimes p_\omega)(f_\rho),\]
 and hence, by \eqref{AutoCorrelationExplicit} and \eqref{PalmDef},
 \[
 \eta^+_{p}(f)  = \bE[(p_\omega \otimes p_\omega)(f_\rho)] = \bE[T\varphi] = (m_G \otimes \nu)(\varphi) = m_G(\check \rho) \cdot  \nu(\cP_\cT f) =  \nu(\cP_\cT f).
  \]
This establishes the formula for $\eta^+_{p}$. In order to obtain the formula for $\eta_p$ it remains to show, in view of \eqref{etap}, that $i(p) = \nu(\cT)$. For any Borel set $V$ we have
\[
M^1_p(V) = \bE[p_\omega(V)] = \bE\left[\sum_{x\in \Lambda_\omega}\chi_V(x)\right] = \bE[\chi_{V.\cT}] = \bP[V.\cT].
\]
If we now choose an identity neighbourhood $V$ in $G$ with $VV^{-1} \subset U$, then
 \[
 i(p) \cdot m_G(V)  = M^1_p(V) = \bP(V.\cT) = m_G(V) \cdot \nu(\cT),
 \]
 and hence $i(p) = \nu(\cT)$ as desired.
 \end{proof}
As a corollary we derive the following formula for the variance.
\begin{corollary}[Palm formula for the variance]\label{VarPalm}
Let $\rho_R := \chi_{B_R} \ast \chi_{B_R}$. Then 
\[
\Var_p(B_R) =  \left(\sum_{\lambda \in \Lambda_\cT} \rho_R(\lambda) \cdot \nu(\cT \cap \lambda^{-1}.\cT)\right) - \nu(\cT)^2 \cdot m_G(\rho_R).
\]
\end{corollary}
\begin{proof} We first observe that
\[
\Var_p(B_R)=\Cov_p(\chi_{B_R} \otimes \chi_{B_R}) = \eta_p(\chi_{B_R} \ast \chi_{B_R}) = \eta_p(\rho_R).
\]
With Proposition \ref{PalmAutocor} we deduce that
\begin{equation}
\Var_p(B_R)=  \nu(\cP_\cT \rho_R) -\nu(\cT)^2 m_G(\rho_R).
\end{equation}
Now for all $\omega \in \cT$ we have
\[
\cP_\cT \rho_R(\omega) = \sum_{\lambda \in \Lambda_\omega}  \rho_R(\lambda) = \sum_{\lambda \in \Lambda_\cT} \rho_R(\lambda) \cdot \chi_{\cT \cap \lambda^{-1}.\cT}(\omega),
\]
hence integrating over $\nu$ yields the desired formula.
\end{proof}

\end{document}